\documentclass[fleqn,usenatbib]{mnras}

\usepackage{graphicx}% Include figure files
\usepackage{dcolumn}% Align table columns on decimal point
\usepackage{bm}% bold math

\usepackage{mathptmx}
\usepackage{mathrsfs}
\usepackage{xcolor}
\usepackage[mathscr]{euscript}
\usepackage{hyperref}
\usepackage{physics}
\usepackage{simplewick}
\usepackage[bottom]{footmisc}
\usepackage{threeparttable}
\usepackage{appendix}
\usepackage{amssymb}
\usepackage{amsmath}
\usepackage{booktabs} % 导入三线表需要的宏包
\usepackage{threeparttable}
\usepackage{multirow}

\newcommand{\be}{\begin{equation}}
\newcommand{\ee}{\end{equation}}

\newcommand{\bs}{\begin{split}}
\newcommand{\es}{\end{split}}
\graphicspath{{./}{figure/}}
\usepackage[T1]{fontenc}
\usepackage{subfigure}

\DeclareRobustCommand{\VAN}[3]{#2}
\let\VANthebibliography\thebibliography
\def\thebibliography{\DeclareRobustCommand{\VAN}[3]{##3}\VANthebibliography}

\title[Wang et al.]{Accretion-modified stellar-mass black hole distribution and milli-Hz gravitational wave backgrounds from galaxy centre}
\author[Wang et al.]{
Mengye Wang,$^{1}$
Yiqiu Ma,$^{2}$\thanks{E-mail: myqphy@hust.edu.cn}
and Qingwen Wu,$^{1}$\thanks{E-mail: qwwu@hust.edu.cn}
\\
$^{1}$Department of Astronomy, School of physics, Huazhong University of Science and Technology, Luoyu Road 1037, Wuhan, China\\
$^{2}$centre for Gravitational Experiment, Hubei Key Laboratory of Gravitation and Quantum Physics, School of physics, Huazhong University of Science and Technology,\\ Luoyu Road 1037, Wuhan, China\\
}
\date{Accepted XXX. Received YYY; in original form ZZZ}
\pubyear{2022}

\begin{document}

\maketitle
\begin{abstract}
Gas accretion of embedded stellar-mass black holes\,(sBHs) or stars in the accretion disk of active galactic nuclei\,(AGNs) will modify the mass distribution of these sBHs and stars, which will also affect the migration of the sBHs/stars. With the introduction of the mass accretion effect, we simulate the evolution of the sBH/star distribution function in a consistent way by extending the Fokker-Planck equation of sBH/star distributions to the mass-varying scenario, and explore the mass distribution of sBHs in the nuclear region of the galaxy centre. We find that the sBHs can grow up to several tens solar mass and form heavier sBH binaries, which will be helpful for us to understand the black-hole mass distribution as observed by the current and future ground-based gravitational wave detectors\,(e.g., LIGO/Virgo, ET and Cosmic Explorer). We further estimate the event rate of extreme mass-ratio inspirals\,(EMRI) for sBH surrounding the massive black hole and calculate the stochastic gravitational wave\,(GW) background of the EMRIs. We find that the background can be detected in future space-borne GW detectors after considering the sBHs embedded in the AGN disk, while the mass accretion has a slight effect on the GW background. 

\end{abstract}
\begin{keywords}
black hole physics-- accretion--  accretion disk -- EMRI -- gravitational wave
\end{keywords}

\section{Introduction}
Stellar-mass binary black holes\,(BBH)  are one of the main target sources for the ground-based gravitational wave\,(GW) detectors such as LIGO, Virgo and KAGRA \citep[][]{LIGO2015,Virgo2015,Kagra2019}. More than 90 stellar-mass BBH merging events have been detected by Advanced LIGO and Advanced Virgo \citep[][]{LIGO2021a,Virgo2015}, which opens the era of GW astronomy. Currently detected events indicate a merger rate $R$ to be about $ 17.3-45\,\rm Gyr^{-3}\,yr^{-1}$ \citep[e.g.,][]{LIGO2021rate,Tiwari2021}. Increasing BBH merger events detected by the GW detectors provides an opportunity to explore the formation channels of these BBHs \citep[][]{Kinugawa2021,Chattopadhyay2022,Ford2021}.  

Two main BBH formation channels are proposed, which are the isolated evolution of massive binaries in galactic fields \citep[e.g.,][]{Belczynski2016,Kinugawa2021} and the dynamical evolution in dense stellar environments \citep[e.g.,][]{Stone2017,Yang2019apj,Chatterjee2017,McKernan2018,Mandel2022,Li2022,Li2022arXiv,Samsing2022Nature}. Strong observational evidence shows that the massive galaxies host massive black holes\,(MBH) in their centre \citep[e.g.,][]{Kormendy2013}, where the MBH mass range from $\sim10^{5-10}M_{\odot}$. The MBH can grow via accreting the surrounding gas and form an accretion disk, which generates an active galactic nuclear (AGN). For an optically thick, geometrically thin standard AGN disk \citep[][]{Shakura1973}, gravitational instability happens at the outer region with several thousand gravitational radii, leading to the star formation via the fragmentation of the gas \citep[e.g.,][]{Sirko2003}. Furthermore, some stars in the nuclear star cluster can also be captured into the accretion disk as a result of momentum and energy loss during the star-disk interaction \citep[e.g.,][]{Artymowicz1993,Vilkoviskij2002,McKernan2012,Kennedy2016,Panamarev2018,Macleod2020}. These disk stars will evolve quickly due to the possible fast accretion of the gas in the AGN accretion disk \citep[e.g., so-called ''accretion-modified star",][]{Wangjianmin2021} and thereby forming compact objects such as stellar-mass black holes\,(sBHs).  These sBHs, together with the sBHs directly captured from the nuclear star cluster by the disk, will migrate inward to the central MBH \citep[e.g.,][]{Vilkoviskij2002,McKernan2012,McKernan2014,Kennedy2016,Panamarev2018,PanZhen2021,PanZhen2022}. The migration of these sBHs and stars will not only be affected by the centre MBH but also by the interaction with the accretion disk \citep[e.g.,][]{Goldreich1979,Goldreich1980,Tanaka2002,Tanaka2004,Fabj2020,Nasim2020,Secunda2019,Secunda2020,Tagawa2020}. The binary system consisting of a sBH and a centre MBH is an important target source of the future space-borne GW detectors \citep[such as LISA/Tianqin/Taiji,][]{Luo2015TianQin,Luo2020,Travis2019LISA}, which is called Extreme-Mass-Ratio-Inspiral\,(EMRI) system.
There are two main channels for the formation of EMRIs, which are called dry EMRI and wet EMRI. The dry EMRIs are mainly produced by multi-body scatterings within the nuclear star cluster and the gravitational capture \citep[such as Hills mechanism, see][]{Hills1988}, while the wet EMRIs are the EMRIs in a gas-rich environment \citep[e.g.,][]{PanZhen2021wet}. For the wet EMRIs, the distributions of nearby stars and sBHs are significantly affected by the dense environments \citep[e.g.,][]{Vilkoviskij2002,McKernan2012,Kennedy2016,Panamarev2018,PanZhen2021wet,PanZhen2022}. 
The wet EMRI rate will be significantly increased compared to the dry EMRI rate \citep[e.g.,][]{PanZhen2021}.

The stars/sBHs embedded in the AGN disk can grow quickly due to the accretion effect. This phenomenon has been considered to explain some unexpected massive binary merger events \citep[e.g., GW190521,][]{Abbott2020}. 
%Furthermore, the migration of the sBHs and stars in the AGN disk also accompanies the mass increase due to the accretion effect. 
The stars and stellar-mass compact objects will also accrete and grow quickly in this dense environment, which will lead to their fast evolution \citep[e.g.,][]{Yangxiaohong2014,McKinney2014,PanZhen2021accretion}.  
%The accretion processes in a dense environment have been discussed in \citet{PanZhen2021accretion} within a general relativistic framework for the sBHs and other compact objects, such as white dwarfs and neutron stars \citep[see also,][]{Yangxiaohong2014,McKinney2014}. 
It should be noted that current simulations for the evolution of the sBH/star distribution function does not consistently involve the accretion effect, that is, the distribution function evolution does not describe the redistribution of the population among different masses. In this work, we re-examine how mass accretion influences the distribution function evolution through more consistent simulations, which are based on the extended Fokker-Planck equations to the mass varying scenario. Moreover, we also phenomenologically take into account the fact that some main sequence stars will collapse into sBHs when their mass exceeds the gravitational limit in our simulations.

The tools for analysing the distribution functions are based on non-equilibrium statistical mechanics, where we consider the stellar objects in the galaxy as interacting gas particles. The sBHs near the galaxy centre interact mutually, with other stars, and also with the MBH of the galaxy centre. From a mean field point of view, these sBHs (together with the main sequence stars and other compact stars) feel a background gravitational potential created collectively by all the matters in the galaxy, upon which these sBHs/stars are also interacting mutually via gravitational scattering. This system is similar to a collection of mutually interacting particles evolving in an external potential, which must satisfy the Fokker-Planck equation \citep[e.g.,][]{Lightman&Shaoiro1977,Shapiro1978,cohn1978,Cohn1979,Pau2011,Stone2017,Stone2018}. This means that, in the phase space, the distribution of these particles will evolve due to the advection and also the interaction-induced diffusion. This non-equilibrium statistical mechanics theory has been applied to analyse the star distribution in a galaxy in the 1970s by \citet{Cohn1979}.

Investigation of the accretion effect on the sBH mass distribution and possible GW sources of BBHs in the galaxy centre could be a topic that relates to the GW observation at different frequency bands (i.e. multi-band GW astronomy). By simulating the evolution of the sBH distribution via these extended Fokker-Planck equations, this work is devoted to analyse the following problems.

(1) The effect of accretion on the sBH mass distribution near the MBH. Future 3rd generation \emph{ground-based} GW detectors are designed to detect nearly all the stellar-mass binary compact objects in the universe \citep[][]{Abbott2017,Maggiore2020}. The effect of accretion studied in this work could be finally tested by the statistics on the properties of these sBH binary systems, which are relevant to their formation history.

(2) The EMRI events, which is an important target source of the proposed \emph{space-borne} gravitational waves detectors such as LISA, Tianqin and Taiji. The accretion process of sBHs in the AGN disk will affect the sBH mass distribution and the EMRI event rate. Therefore, it could potentially affect the observation of the GW background emitted from the EMRIs in the gas-rich environment by future space-borne GW detectors. In this work, the effect of the accretion on the EMRI rate and the stochastic GW background\,(GWB) from the EMRIs will be discussed.

%(3) Tidal disruption event, which strongly depends on the star distribution near the centre MBH, \textbf{typically observed in the quiet galaxies}. In these cases, the accretion process will not affect the TDE event. However, recent observations show that XXX. 
This paper is organized as follows. We present the extended Fokker-Plank equations by considering the mass variation effect in Section~\ref{sec:2}; The simulation setup and simulation results are shown in Sections~\ref{sec:3} and \ref{sec:4}, respectively; The effect of accretion on the distribution of sBHs, together with its possible connection to the gravitational wave detection using ground-based detectors is investigated in Section~\ref{sec:5}; In Sections~\ref{sec:6} and \ref{sec:7}, taking into account the accretion effect, we study the EMRI event rate, the stochastic GWB from these EMRIs, and the observational consequence of the space-borne detectors. Discussions and conclusions are given in Section~\ref{sec:8}.

%\begin{figure}
%\centering
%\includegraphics[scale=0.18]{BH_distribution.jpeg}
%\caption{Black hole mass distribution function. The blue histogram shows the distribution of the mass of the black holes detected by the current LIGO detector, before the coalescence. The red line is our simulation result using the modified Fokker-Planck equation, which appears to fits well with the observational result.}
%\label{fig:1}
%\end{figure}
\section{Fokker-Planck equation with varying mass}\label{sec:2}
\subsection{Extended Fokker-Planck equation}

To include the mass variation effect in the evolution of the distribution function, we extend the original phase space $(E, J)$ with an additional dimension to the phase space $(E, J, m)$, where the parameters $E$ and $J$ are the orbital energy per unit mass\,(specific orbital energy) and the orbital angular momentum per unit mass\,(specific angular momentum). The additional dimension counts the mass variation. Therefore, the new distribution function in the phase space $f(E,J,m)$ is proportional to ${\rm d}N/({\rm d} E{\rm d}J{\rm d}m)$, where ${\rm d}N$ is the number of particles within $(E, E+{\rm d}E), (J, J+{\rm d}J)$ and $(m, m+{\rm d}m)$. The Boltzmann equation in this new phase space is:
\begin{equation}\label{eq:boltzmann}
\frac{\partial f(E,J, m)}{\partial t}+\mathbf{\nabla}\cdot[f( E,J, m)\mathbf{v}]=\text{collision terms},
\end{equation}
where $\mathbf{\nabla}=(\partial_E,\partial_J,\partial_m)$. We also have $\mathbf{v}=(\dot{E},\dot{J},\dot{m})$, which is only affected by the mean-field background gravitational potential of the galaxy, and the accretion rate of the stars and sBHs. %The collision term on the right-hand-side is due to the gravitational scattering of the particles, which can be treated as a stochastic perturbation on the velocity $\bar{\mathbf{v}}+\delta\mathbf{v}$. In this way, we can rewrite the equation~\eqref{eq:boltzmann} as:
%\be\label{eq:boltzmann2}
%\frac{\partial f(J, E, M)}{\partial t}+\nabla\cdot(f(J, E, M)\mathbf{\bar v})=-\nabla\cdot(f(J, E, M)\delta\mathbf{v}).
%\ee
%We assume that (1) $\delta v_3(t)=0$ which means the mass change is deterministic since the accretion is steady when the BHs/stars migrate in the disk region where the gas supply is sufficient, (2) the perturbation $\delta\mathbf{v}$ of $E$ and $J$ satisfies an unbiased Gaussian distribution:
%\be
%\langle\delta v_i(t)\delta v_j(t')\rangle
%=2D_{ij}\delta(t-t'), with 
%\ee
Compared to the previous result obtained in \citet{Lightman&Shaoiro1977,Shapiro1978,Cohn1979,Merritt2013,Stone2018,PanZhen2021,Broggi2022}, the resulting Fokker-Planck equation will have one more term that relates to the mass variation, while other terms would be the same or a straightforward extension of the equations in \citet{Cohn1979}. Following the conventions of \citet{Cohn1979}, we re-parametrize the phase space as $(E,\mathscr{R})$ with $\mathscr{R}\equiv J^2/J_c^2$, where $J_c=GM_\bullet\sqrt{1/2E}$ is the specific orbital angular momentum of a sBH with specific energy $E$ on a circular orbit and the $M_\bullet$ is the mass of centre massive black hole\,(MBH). Given initial distributions of stars and sBHs $f_{s/\rm sBH} (t = 0, E, \mathscr{R} ,m)$, their evolution is governed by the orbit-averaged Fokker-Planck equation
\be\label{eq:FP_eqn}
\mathcal{C}\frac{\partial f}{\partial t}=-\frac{\partial}{\partial E}F_E-\frac{\partial }{\partial \mathscr{R}}F_\mathscr{R}-\mathcal{C}\frac{\partial}{\partial m}(f\dot{m}),
\ee
where $\mathcal{C}$ is the weighting function defined through $\mathcal{C}(E,\mathscr{R})\equiv 4\pi^2 P(E,\mathscr{R}) J_c^2(E)$, $P(E,\mathscr{R})$ is the orbital period and $F_{E,\mathscr{R}}$ is the flux given by \citep[][]{Shapiro1978,cohn1978}:
\be \label{eq:FER}
\begin{split}
&F_E=-\left(D_{EE}\frac{\partial f}{\partial E}+D_{E\mathscr{R}}\frac{\partial f}{\partial \mathscr{R}}+D_E f\right),\\
&F_\mathscr{R}=-\left(D_{\mathscr{RR}}\frac{\partial f}{\partial \mathscr{R}}+D_{E\mathscr{R}}\frac{\partial f}{\partial E}+D_\mathscr{R} f\right).
\end{split}
\ee
The terms with diffusion coefficients $(D_{EE},D_{E\mathscr{R}},D_{\mathscr{RR}})$ represent
the relaxation of the particles in the phase space due to the energy and angular momentum exchange induced by gravitational interaction. We assume that the accretion process is not stochastic and the mass increase of the sBHs and stars is steady, therefore there is no direct participation of the accretion process in the diffusion. The $(D_E, D_\mathscr{R})$ terms, together with the $m$-derivative term in equation~\eqref{eq:FP_eqn} are the advection coefficients, where the $(D_E, D_\mathscr{R})$ terms describe the advection due to the energy and angular momentum increment in the process of two-body scattering. The $m$-derivative term in equation~\eqref{eq:FP_eqn} describes the advection in the $m$-direction through accretion. 
However, mass accretion indirectly affects the diffusion process in the $(E,\mathscr{R})$ directions. This is because the gravitational two-body scattering, which is the source of the diffusion process, depends on the mass of the encountering particles. This can be naturally extended in the following way. For example, the diffusion in the energy direction now can be rewritten as:
\be
\begin{split}
&D_{EE}(E,\mathscr{R},m)= \\
&\frac{8\pi^2}{3}J_c^2\int dm'\int^{r_+}_{r_-}\frac{dr}{v_r}v^2\left[F_0(m',E,r)+F_2(m',E,r)\right],\\
&{\rm with}\\
&F_0(m,E,r)=(4\pi)^2 {Gm}^2\ln{\Lambda}\int^E_{-\infty}dE'\bar f(m, E'),\\
&F_2(m,E,r)=(4\pi)^2 {Gm}^2\ln{\Lambda}\int_E^{\phi(r)}dE'\left(\frac{\phi-E'}{\phi-E}\right)^{3/2}\bar f(m, E'),
\end{split}
\ee
where $r_+/r_-$ is the orbiters' apocentre/pericenter radius, $v_r$ is the radial velocity, $\bar f(m,E)=\int d\mathscr{R} f(m,E,\mathscr{R})$, $\ln \Lambda=10$ is the Coulomb’s logarithm \citep[][]{Binney2008,Pau2018} and $\phi (r)$ is the gravitational potential defined in equation \eqref{eq:phi}.
The other diffusion coefficients can be extended similarly, as listed in Appendix \ref{appendix}.  In the later simulation, the particles will be grouped in terms of mass to discretize these differential equations.

When the MBH has an accretion disk, the above diffusion and advection coefficients will be affected by the accretion disk. The star-disk interaction can excite the density wave \citep[][]{Goldreich1979,Goldreich1980,Tanaka2002,Tanaka2004}, which will speed up the migration of stars/sBHs. This migration process is driven by the star-disk interaction, thereby being a deterministic process if the accretion disk has a steady structure. Therefore, the interaction between the accretion disk and the stars/sBHs will mostly affect the advection term (see Section \ref{sec:3} for details), while the diffusion terms are still determined by the random gravitational scattering between the stars/sBHs \citep[see also,][]{PanZhen2021}. 

\subsection{sBH/star accretion in the AGN disk}
Generally speaking, the accretion process of sBHs in AGN disk is affected by many factors and therefore is quite complicated \citep[e.g.,][]{Wangjianmin2021}. The widely adopted Bondi accretion model is over-simplified and many physical effects should be taken into account, such as radiative feedback, vertical stratification, shear viscosity, tidal effects and gap opening, etc. \citet{Yangxiaohong2014} and \citet{McKinney2014} have studied the two- and three-dimensional general relativistic radiation magnetohydrodynamical simulation of super-Eddington accretion.
Moreover, \citet{PanZhen2021accretion} studied the supercritical accretion of stellar-mass compact objects in the AGN disk using a general relativistic framework, which considered both the inflow and outflow. Their results suggested that the sBHs' accretion rate is roughly $\sim 10 \,\dot{M}_{\rm Edd}$ even if the inflow rate at the outer edge of the sBH accretion disk is very high. Considering the disk wind under super-Eddington accretion, it is found that the intrinsic accretion rate is limited to several times of Eddington rate \citep[e.g.,][]{Dotan2011,Gu2012,Fengjunjie2019}. In the outer region, the disk is unstable so that it could be fragmentation and collapse into clumps \citep[][]{Sirko2003,Durisen2007,Andera2022}. In this way, there may not be enough gas for accretion in the outer unstable region. In this work, we set the accretion rate of sBHs in AGN disk as a free parameter, which ranges from several to several tens Eddington accretion rate ($\dot{M}_{\rm Edd}$) in the inner stable region and equal to $\dot{M}_{\rm Edd}$ if the sBH stay in the outer unstable region.

The main sequence stars usually have a higher accretion rate than the sBH. Similarly, it is still a lack of a sufficiently clear understanding of the detailed accretion process owing to the complexities of stellar feedback (radiation and winds) and its impacts on the disk structure \citep[][]{Matto2021,Dittmann2021,Jermyn2021}. In the stable region of the AGN disk, to avoid numerous uncertainties in the modeling of those effects, we use the following accretion model fit by the  top-heavy stellar mass distribution, which is inferred from the universally high abundance ratio of [Fe/Mg]
\citep{Toyouchi2022},
\be\label{eq:star_accretion}
\dot{m}_s=\dot{M}_{0.1}\left(\frac{m_s}{0.1M_{\odot}}\right)^2\left(1+\frac{m_s}{M_c}\right)^{\alpha-2},
\ee
where $M_c\, ,\alpha$ and $\dot M_{0.1}$ are three tunable parameters  \citep[see Table 1 in][]{Toyouchi2022}. Here, we adopt the model B in \citet{Toyouchi2022}, where the parameters are $\dot M_{0.1}=2.3\times 10^{-8}\,M_\odot \,\rm yr^{-1}$, $M_c=9.4\,M_\odot$, $\alpha=-0.5$ and the maximum star mass $M_{\rm max}=300\,M_\odot$. We also assume $\dot{m}_s=\dot{M}_{\rm Edd}$ in the unstable region of the AGN disk. With increasing masses, some main sequence stars will collapse and become sBHs, if the orbits of these stars are lying on the accretion disk of the centre MBH.
It means that there will be a source term in equation~\eqref{eq:FP_eqn} for the sBH distribution on the accretion disk, as we shall present more carefully in the next section.

The gas in the accretion process is supplied by the dense disk environment in the AGNs. However, the active galaxies may have duty cycles \citep[e.g.,][]{Shulevski2015,Turner2018}, that is, the MBH is only active in some stages, where the accretion disk of the central MBH will diminish during its quiet stage. In the following simulation, we assume that the mass of stars and sBHs remains unchanged during the quiet stage, and turn on the $\dot m$ in the Fokker-Planck equation only when the galaxy is in its active stage.

\section{Simulation setup}\label{sec:3}
We consider the following two-stage physical scenario. Given the initial distribution function of the star and sBH in the nuclear region of galaxies, we first simulate the star/sBH distribution function for 5\,Gyr in the quiet stage of the galaxy, without the participation of the centre MBH's accretion disk. In this quiet stage, the evolution is spherically symmetric. After that, we turn on the accretion disk when the galaxy enters its active stage. In this active stage, the plane of the accretion disk corresponding to the latitude angle $\theta=\pi/2$ becomes specific, and the spherical symmetry of the evolution will break. The general framework of the numerical method is based on the work of \citet{PanZhen2021,Broggi2022}.

\subsection{Quiet stage evolution}
In the quiet stage, the standard Fokker-Planck equations are given in the literature \citep[e.g.,][]{cohn1978,Cohn1979}, i.e., set the accretion rate $\dot{m}_{s/\rm sBH}=0$ of stars/sBHs in equation~\eqref{eq:FP_eqn}. We set the initial and boundary condition following Tremaine's MBH+stellar cluster model \citep[][]{Tremaine1994}.

The stars in the stellar cluster are assumed to have a single mass component with $m_{s}=1M_\odot$, while the mass of sBH in the stellar cluster is distributed within $m_{\rm sBH}=5-15M_\odot$. The total mass of stars is fixed to be $M_s=20M_\bullet$ \citep[see also,][]{PanZhen2021,Broggi2022}, where $M_\bullet$ is the mass of the central MBH and the total mass of sBHs is denoted as $M_{\rm sBH}$. In Tremaine's cluster model, the initial number densities of stars and sBHs with different masses are given by \citep[see also,][]{Binney2008}:
\be \label{number_density}
\begin{split}
&n_{s}(m_s,r)=\frac{M_s}{m_s}\frac{3-\gamma}{4\pi}\frac{r_a}{r^\gamma (r+r_a)^{4-\gamma}}\delta(m_s-M_\odot), \\
&n_{\rm sBH}(m_{\rm sBH},r)=\varphi (m_{\rm sBH}) \times n_s(r),
\end{split}
\ee
where $r_a=4GM_\bullet/\sigma_*^2\equiv 4r_h$ is the radius of density transition\,($r_h$ is the influential radius of the MBH with mass $M_{\bullet}$), $\sigma_*$ is the stellar velocity dispersion given by $M_\bullet-\sigma_\star$ relation \citep[][]{Tremaine2002,Kayhan2009},
\be \label{M_sigma_relation}
M_\bullet=1.53\times 10^6M_\odot \left(\frac{\sigma_*}{70 {\rm km}/s}\right)^{4.24}
\ee
and $\gamma\sim 1.2-1.8$ is the density scaling power index \citep[][]{Tremaine1994,Binney2008}. The $\varphi (m_{\rm sBH})\propto m_{\rm sBH}^{-2.35}$ \citep{Salpeter1955} is the relative abundance of sBHs with different mass and $n_s(r)\equiv\int n_s(m_s,r)d\,m_s$ is the total number density of the stars. The total relative abundance of sBHs is defined as
\be
\varphi=\int_{5M_\odot}^{\rm 15M_\odot}\varphi (m_{\rm sBH})\, {\rm d}m_{\rm sBH},
\ee
of which the value is assumed to be  $0.001-0.002$ in \cite{PanZhen2021}.

With the above density profiles, the collective gravitational potential background can be derived as:
\be \label{eq:phi}
\phi (r)=\frac{GM_\bullet}{r} +\frac{G(M_{s}+M_{\rm sBH})}{r_a}\frac{1}{2-\gamma}\left[1-\left(\frac{r}{r+r_a}\right)^{2-\gamma}\right].
\ee
In this case, the initial distribution function in the $(E,\mathscr{R},m)$-phase space is \citep[][]{Tremaine1994,Binney2008}%only depend on the energy $E$ and the space number density of stars/sBHs by
\be
f_{s/\rm sBH}(t=0,E,\mathscr{R},m_{s/\rm sBH})=\frac{\sqrt{2}}{(2\pi)^2}\frac{{\rm d}}{{\rm d}E}\int^E_0\frac{{\rm d}n_{s/\rm sBH}}{{\rm d} \phi}\frac{{\rm d} \phi}{\sqrt{E-\phi}},
\ee
where $n_{s/\rm sBH}$ is the number density of stars/sBHs (see equation \eqref{number_density}).

The boundary conditions are set as follows, which is widely adopted in literature  \citep[e.g.,][]{,cohn1978,Cohn1979,Merritt2013,PanZhen2021,Broggi2022}:

(1) At $E=0$ where the stars and sBHs are distributed far away from the central MBH, their long relaxation time allows us to assume the distribution function to be independent of time and equal to the initial distributions. 

(2) The boundary $\mathscr{R}=1(\text{or equivalently}\,\,J=J_c(E))$ corresponds to circular orbits, which defines the edge of the phase space in the $\mathscr{R}$-direction. Therefore, the flux in the $R$-direction should vanish for both stars and sBHs, i.e. $F^{s/\rm sBH}_\mathscr{R}|_{\mathscr{R}=1}=0$.

(3) At loss-cone boundary $\mathscr{R}=\mathscr{R}_{\rm lc}(E)$, the behaviour of $f_{s/\rm sBH}$ has been derived by Cohn and Kulsrud in \citet{Cohn1979} and \citet{Merritt2013} as 
\be 
f_{s/\rm sBH}(\mathscr{R})\approx f_{s/\rm sBH}(\mathscr{R}_{\rm lc})\left(1+\frac{{\rm ln}(\mathscr{R}/\mathscr{R}_{\rm lc})}{{\rm ln}(\mathscr{R}_{\rm lc}/\mathscr{R}_0)} \right),\quad \mathscr{R}\rightarrow \mathscr{R}_{\rm lc},
\ee
where $R_0$ is given by the following approximate relation 
\be
\mathscr{R}_0\approx \mathscr{R}_{\rm lc}\ {\rm exp}\left(-\sqrt[4]{q^4+q^2}\right),
\ee
 with
\be
q=\frac{1}{4\pi^2J_c^2\mathscr{R}_{\rm lc}} {\rm lim}_{\mathscr{R}\rightarrow \mathscr{R}_{\rm lc}}\frac{D_{ \mathscr{RR}}}{\mathscr{R}}.
\ee
The limit at a given value of $E$ is numerically performed by evaluating the quantity $q$ at the first grid point above the loss-cone curve.

\subsection{Disk-star/sBH Interactions}
In the active stage, we adopt the standard thin $\alpha-$disk model for the stable region of accretion disk \citep[][]{Shakura1973}. The outer parts of the accretion disk will be prone to unstable if Toomre’s stability parameter $Q$ satisfies: 
\be
Q\equiv\frac{c_s\Omega}{\pi G\Sigma}\approx\frac{\Omega^2}{2\pi G\rho}<1.
\ee
Here, we accept the SG-disk model which maintains a minimum value of the Toomre's parameter $Q=1$ by external feed-back heating in the outer region \citep{Sirko2003}, which basically means that the disk outer boundary will extend further.
 %a larger accretion rate at the disk's outer edge than expected for the usual interstellar medium.
The sBH-disk interactions have been discussed in many papers \citep[see][for reviews]{Kley2012,papaloizou2021,Paardekooper2022}, which are briefly summarised as follows. %In the outer region, we assume some external feeding back mechanism heats the disk and maintains a minimum value of the Toomre’s parameter as $Q=1$ \citep{Sirko2003}, which is called SG-disk. With the presence of an accretion disk, the sBH-disk interaction will reshape the sBHs/stars' distribution. In this work, we consider two sBH-disk interaction regimes including density wave and headwind, which have been discussed by \citet{PanZhen2021}. Here, we briefly summarize these two interactions.}

\textbf{Type I migration}--- The periodic orbital motion of sBHs around the MBH excites the density waves, which consist of three components: regular density waves excited by the circular orbit, eccentricity waves excited by the non-circular orbit and bending waves excited by the orbit normal to the disk \citep[e.g.,][]{Goldreich1978,Goldreich1980,Tanaka2002,Tanaka2004}. The regular density waves exert a (type-I) migration torque on the sBH and drive its migration in the radial direction with the timescale $\tau_{\rm mig,I}$; the eccentricity and bending density waves damp the orbit eccentricity and the inclination angle to the disk plane on the timescale $\tau_{\rm wave}$. The type-I migration torque can be formulated as \citep[][]{Tanaka2002,Tanaka2004}
\begin{equation}
    \dot{J}_{\rm mig,I}=C_I \frac{m_{\rm sBH}}{M}\frac{\Sigma}{M}\frac{r^4\Omega^2}{h^2},
\end{equation}
where $M=M(<r)$ is the total mass consisting of the MBH, stars, sBHs and the disk gas within the radius $r$, $C_I=-0.85+{\rm d} {\,\rm log\,} \sigma_*/{\rm d} {\,\log{r}}+0.9{\rm d}{\,\rm log\,}T_c/{\rm d}{\,\log{r}} $, $\sigma_*(r)$, $T_c(r)$, $h(r)=H(r)/r$, $\Omega(r)$ are the disk surface density, the disk middle plane temperature, the disk aspect ratio ($H(r)$ is the thickness of the disk at radius $r$) and the sBH angular velocity, respectively \citep[][]{Paardekooper2011}. The damping timescale of the orbit eccentricity and the inclination is 
\begin{equation} \label{eq:tau_wave}
    \tau_{\rm wave}\approx h^2\tau_{\rm mig,I}=h^2\frac{J}{\dot{J}_{\rm mig,I}}\sim \frac{M}{m_{\rm sBH}}\frac{M}{\Sigma r^2}\frac{h^4}{\Omega},
\end{equation}
where $J=r^2\Omega$ is the specific angular momentum of sBH.

\textbf{Type II migration}---The type-I migration is replaced by type-II if the sBH is so massive that a gap in the disk opens up. When the gap opening occurs, the motion of the sBH is locked to the viscous evolution of the disk, hence, the type-II migration torque is \citep[e.g.,][]{Lin1986,Ward1997,Syer1995} 
\begin{equation}
    \dot{J}_{\rm mig,II}=-\frac{2\pi r^2\Sigma}{m_{\rm sBH}}r\Omega |v_{{\rm gas},r}|,
\end{equation}
where $v_{{\rm gas},r}=-\dot{M}_\bullet/(2\pi r\Sigma)$ is the gas inflow velocity .
%\textbf{Comment: Are you sure this is the correct formula? since $Mv\Omega=Mr^2\Omega$ has no direct physical meaning if $M$ is the total mass of MBH, stars, sBH and the disk within $r$. I initially thought it would represent the angular momentum at radius $r$.??\R{WMY}: Here, $M$ should be the mass of the disk. It means that the angular momentum transfer from the disk to sBH.  \textbf{Inconsistent definition! see your definition below equation~13.} \R{I mean the $M$ in $\frac{1}{m_{\rm sBH}}\frac{d}{dt} (Mr^2\Omega)$ should be disk mass, or in another word,  } }

\textbf{Head wind}---For a sBH embedded in the gas disk, surrounding gas in its gravitational influence sphere flows towards it. Considering the differential rotation of the disk, the inflow gas generally carries nonzero angular momentum relative to the sBH. Therefore the inflow tends to form a certain local disk or bulge profile around the sBH. Radiative feedback and magnetic effects drive a major part of the captured material to escape in the form of outflow, therefore only the remaining part is accreted by the sBH \citep[][]{Yangxiaohong2014,McKinney2014}. These accreted materials from the headwind can transfer angular momentum to the orbital motion of the sBH, hence exerting a specific torque given by:
\begin{equation}
    \dot{J}_{\rm wind}=-\frac{r\delta v_\phi\dot{m}_{\rm gas}}{m_{\rm sBH}},
\end{equation}
where $\delta v_\phi=v_{\phi, \rm gas}-v_{\phi,\rm sBH}$ is the head wind speed and $\dot{m}_{\rm gas}$ is gas capture rate \citep[see][for detailed calculation]{PanZhen2021}.

In a word, the sBH migration timescale in the AGN disk and the nuclear star cluster are \citep[see also,][]{PanZhen2021}
\begin{equation}
    \tau_{\rm mig}^{\rm d}=\frac{J}{|\dot{J}_{\rm mig,I/II}+\dot{J}_{\rm wind}+\dot{J}_{\rm gw}|}, \quad \tau_{\rm mig}^{\rm c}=\frac{J}{|\dot{J}_{\rm mig,I}+\dot{J}_{\rm gw}|},
\end{equation}
where the specific torque by GW emissions $\dot{J}_{\rm gw}$ is given by
\begin{equation}
    \dot{J}_{\rm gw}=-\frac{32}{5c^5}\frac{m_{\rm sBH}}{M} \left(\frac{GM}{r}\right)^{7/2}.
\end{equation}
When a gap is opened, we set the headwind torque $\dot{J}_{\rm wind}=0$.

The evolution of those stars/sBHs with the orbit plane parallel to the accretion disk will be significantly different from the rest. Here we use the following ansatz to study the distribution function evolution of these stars/sBHs on the disk.

\subsection{Active stage evolution}
To distinguish those stars/sBHs on the disk plane, we name these stars/sBHs as "disk stars/sBHs" in the following and denote their distribution function as $g_{s/\rm sBH}(t, E, \mathscr{R}, m)$, while we still use $f_{s/\rm sBH}$ to describe the distribution of the star/sBHs with orbits out of the disk plane\,(in the following these stars/sBHs will be named as ``cluster stars/sBHs''). Before we write down the evolution equations for the cluster/disk stars/sBHs distributions, several important points need to be noticed:

(1) The orbit of the disk stars/sBHs will be quickly circularized due to their interaction with the accretion disk (see Eq.\eqref{eq:tau_wave}). This indicates that we can make the approximation $\mathscr{R}\approx 1$ in the $g_{s/\rm sBH}(t, E, \mathscr{R}, m)$, and therefore we reduce the $\mathscr{R}-$direction in the phase space and define:
\be
g_{s/\rm sBH}(t, E, m)=\int_0^1 g_{s/\rm sBH}(t, E, \mathscr{R}, m)\,{\rm d}\mathscr{R}.
\ee
Moreover, the advection term $D_E$ in the $E$-direction overwhelms the diffusion term $D_{EE}$ for the disk stars/sBHs because the star-disk interaction mainly affects the advection term.
Therefore, we can approximate $D_{EE}=0$ in the evolution equation of $g_{s/\rm sBH}(t, E, m)$.

(2) The exchange of stars/sBHs between the nuclear star cluster and the disk involves two processes: the capture of the cluster stars/sBHs by the disk through the damping of the orbital inclination angle, and the scattering of the stars/sBHs from the disk to the cluster. The timescale of these two processes is much shorter than the migration timescale, especially for orbiters with small inclination angles. Therefore, a local equilibrium can be approximately established, which leads to the phenomenological source term in the Fokker-Planck equation given as $S_{s/\rm sBH}= \mu_{s/\rm sBH}f_{s/\rm sBH}/\tau^{\rm d}_{\rm mig}$ \citep[see also,][]{Vilkoviskij2002,Kennedy2016,Panamarev2018,PanZhen2021}, where the $f_{s/\rm sBH}$ is the stars/sBHs distribution function in the nuclear star cluster. The net capture rate will be determined by parameter $\mu_{s/\rm sBH}$.
Here, we set $\mu_{\rm sBH}=\mu_s m_{\rm sBH}/m_s$ and $\mu_s $ is an adjustable parameter \citep[see also][]{PanZhen2021}. It is clear to see that the larger cluster population $f_{s/\rm sBH}$ and stronger migration\,(shorter $\tau^d_{\rm mig}$) means a larger capture effect.

(3) As mentioned in the previous section, the collapse of the massive stars to the sBHs is a complicated process that depends on many physical factors, this process is phenomenologically treated with a lifetime $T_s$ of stars in disk.  Note that we assume a homogeneous mass distribution of the stars ($m_s=M_\odot$) in the nuclear star cluster where the accretion is absent, therefore $f_s(m_s,E)$ in the equation~\eqref{eq:FP_disk} can be written as $f_s(m_s,E)=f_s(E)\delta (m_s-M_\odot)$.

%The interaction between the disk and the orbiting stars/sBHs manifests in the modification of advection coefficients since the interaction is mostly a deterministic process, as carefully discussed in [Pan]. We summarise their main results in the Appendix. \textbf{Moreover, near the outer edge of the disk with active star formation, the disk density drops significantly, therefore the accretion rate in this region will also decrease.}

\begin{figure}
\centering
\includegraphics[scale=0.2]{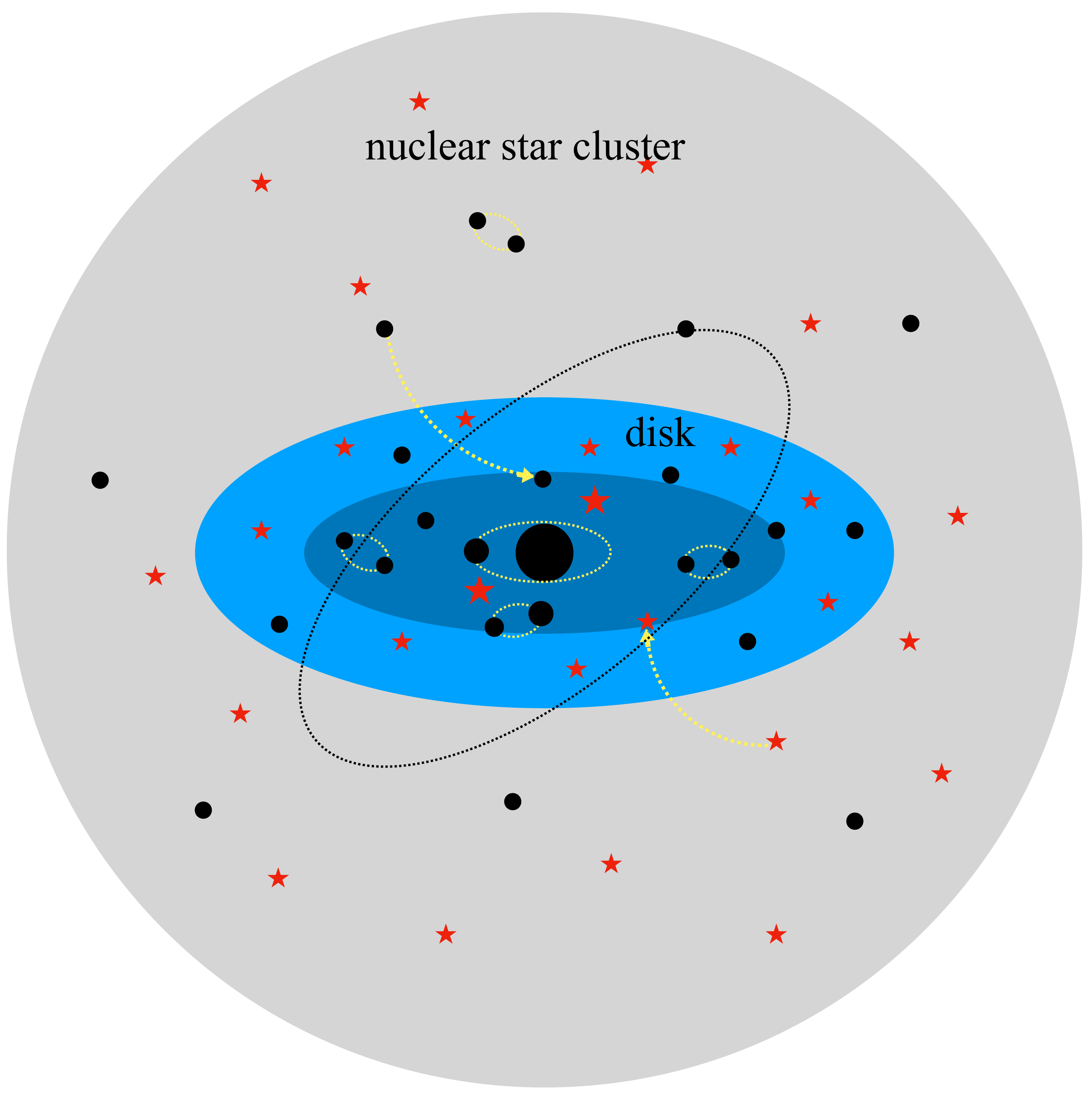}
\caption{Schematic picture of the physical scenario in our work. A MBH is located at the galaxy centre, with an surrounding accretion disk and a population of stars/sBHs. The disk can capture the stars/sBHs from the nuclear star cluster and increase the EMRI rate by the star/sBH-disk interaction. For stars/sBHs embedded in the disk, they will grow by accreting the disk gas. Meanwhile, under the influence of the sBH-disk interaction, sBHs in the disk will can form binaries and merge, which contributes to the gravitational wave sources. In the main text, the Fokker-Planck equation is adopted to describe the evolution of the mass distribution function for sBHs and stars.}
\label{fig:1}
\end{figure}

%Sketch of the AGN disk and the stellar-mass BH orbit indicating the parameters used in the calculation.

Now we are ready to write down the evolution equations in the active stage:
\be\label{eq:FP_disk}
\begin{split}
&\mathcal{C}\frac{\partial f_{\rm sBH}}{\partial t}=-\frac{\partial F^{\rm sBH}_E}{\partial E}-\frac{\partial F^{\rm sBH}_\mathscr{R}}{\partial \mathscr{R}}
-\mathcal{C}\mu_{\rm sBH}\frac{f_{\rm sBH}}{\tau^{\rm d,sBH}_{\rm mig}},\\
&\mathcal{C}\frac{\partial f_s}{\partial t}=-\frac{\partial}{\partial E}F^{s}_E-\frac{\partial }{\partial \mathscr{R}}F^{s}_\mathscr{R}-\mathcal{C}\mu_s\frac{f_s}{\tau^{{\rm d},s}_{\rm mig}},\\
&\frac{\partial g_{\rm sBH}}{\partial t}\approx \frac{1}{\mathcal{C}} \frac{\partial (D_E^{\rm d, sBH} g_{\rm sBH})}{\partial E}-\frac{\partial (g_{\rm sBH}\dot{m}_{\rm sBH})}{\partial m}+\\
&\quad\quad\quad\quad\frac{g_{s}(\epsilon m_{\rm sBH})}{T_s}
+\mu_{\rm sBH}\frac{\bar{f}_{\rm sBH}}{\tau^{\rm d,sBH}_{\rm mig}},\\
&\frac{\partial g_{s}}{\partial t}\approx \frac{1}{\mathcal{C}}\frac{\partial (D^{{\rm d},s}_E g_s)}{\partial E }-\frac{\partial (g_{s}\dot{m}_{s})}{\partial m}-\frac{g_{s}( m_s)}{T_{s}}
+\mu_s\frac{\bar{f}_{s}}{\tau^{{\rm d,}s}_{\rm mig}},\\
\end{split}
\ee
where $\bar{f}_s=\int^1_0 {\rm d}\,\mathscr{R} f_s$, $\bar{f}_{\rm sBH}=\int^1_0 {\rm d}\,\mathscr{R} f_{\rm sBH}$ and $T_s$ is the lifetime of stars in disk. In addition, $F_E$, $F_\mathscr{R}$ is defined in equation~\eqref{eq:FER}, with the advection coefficients modified as
\be
\begin{split}
&D_E^{i,j}\rightarrow D_E^j-\mathcal{C}\frac{E}{\tau_{\rm mig}^{i,j}},\\
&D_\mathscr{R}^{i,\rm c}\rightarrow D_\mathscr{R}^{j}-\mathcal{C}\frac{1-\mathscr{R}}{\tau_{\rm mig}^{{\rm c},j}},
\end{split}
\ee
where $i=\{c, d\}$ and $j=\{\rm star, sBH\}$. The superscript "c" and "d" mean "cluster" and "disk", respectively.  Note that we set $\epsilon=10$ since roughly 90\% of the mass of the progenitor star for $50\,M_\odot<m_s<150\,M_\odot$ will be lost during the supernova explosion \citep[see][]{Nomoto2013}. We also assume the lifetime $T_s$ of stars with mass in range of $[50\,M_\odot,150\,M_\odot]$ is $5\,\rm Myr$ \citep[e.g.,][]{Toyouchi2022}. For stars with $m_s>150M_\odot$, they undergo thermonuclear explosions triggered by pair-creation instability. Such stars are completely disrupted without forming a sBH \citep[e.g.,][]{Umeda2002,Heger2002,Nomoto2013}. We don't consider the gravitational collapse of the stars with mass $m_s<50\,M_\odot$ since their lifetime is much longer than $5\,\rm Myr$. %We neglect the supernova remnants of stars with mass $m_s<50M_\odot$ since their lifetime is further than $5\rm Myr$. 

The initial condition for solving the above distribution evolution is the results of the 5\,Gyr spherically symmetric evolution when the disk is absent. If we assume that the solid angle spanned by the disk is around 1\%, then the initial distribution function for cluster stars/sBHs is given by 
$f_{s/\rm sBH}(t^{\rm active}_{\rm ini})=99\%\times f_{s/\rm sBH}(t^{\rm quiet}_{\rm end})$, while $g_{s/\rm sBH}(t^{\rm active}_{\rm ini},E,m_{s\rm /sBH})=1\%\times \int {\rm d}\,\mathscr{R}f_{s/\rm sBH}(t^{\rm quiet}_{\rm end})$ for disk stars/sBHs, which is similar to the settings in\,\citet{PanZhen2022}.

The boundary conditions for the distribution of the cluster stars/sBHs are the same as the quiet stage at the boundary of $E=0$ and $\mathcal{R}=1$. At the loss cone boundary in the active stage, it can be set $F_R=0$ since the fast eccentricity damping by density waves drives stars/sBHs away from the loss cone \citep[see also][]{PanZhen2021}.
For the disk stars/sBHs, only two boundary conditions are required, which is 
\be
\begin{split}
&{g_i(t,E)|_{E\rightarrow 0}=g_i(t=0,E)|_{E\rightarrow 0}}, \\
&\left.\frac{\partial [\dot m_i g_i(t,E,m_i)]}{\partial m_i}\right|_{m_i=m_{i,\rm min}}=\frac{\dot m_i g_i(t,E,m_{i,\rm min})}{\Delta m_i},
\end{split}
\ee
where $m_{i,\rm min}$ is the minimum mass of sBH/star, $\Delta m_i$ is the mass grid in $\log$-scale and $i=\{\rm star,\,sBH\}$. For the sBHs, the $\dot m_{\rm sBH}$ takes the value of $1\,\dot M_{\rm Edd}$ in the outer disk region and $2-50\,\dot M_{\rm Edd}$ in the inner disk region. For the stars, we set the $\dot m_{s}$ takes the value of $1\,\dot M_{\rm Edd}$ in the outer disk region and follows equation\,\eqref{eq:star_accretion} in the inner disk region. The first boundary condition comes from the approximation that the migration rate when $E\rightarrow 0$ is very low.

\section{Simulation Results}\label{sec:4}
%There are two components in the nuclear star cluster: the stars with mass $m_s=1M_\odot$ and the sBHs with masses $m_{\rm sBH}\in [5,15]\,M_\odot$ \citep[see][for a review]{Nomoto2013}, and also two components in the AGN disk: the stars and sBHs with mass determined by their accretion rate and the accretion time.
Following the above procedure, we consider a model consisting of stars/sBHs orbiting around a MBH ($M_\bullet\in[10^5,10^{10}]\,M_\odot$), where part of the stars/sBHs embedded in the accretion disk while others still stay in spherical star cluster. We set the initial condition to be the stars with a monochromatic mass distribution at $m_s=1M_\odot$ and the sBHs with masses of $m_{\rm sBH}\in [5,15]\,M_\odot$ if they stay in spherical star cluster \citep[see][for a review]{Nomoto2013}. It should be noted that, in the simulation of the Fokker-Planck equation, the loss rate enhancement of stars using the complete mass distribution function compared to that of using the monochromatic mass function is in the range of $1-2$
\citep[see][]{Magorrain1999MNRAS,Kennedy2016}. Thus, for numerical simplicity, we adopt a monochromatic mass function for cluster stars, which will not affect our main conclusion \citep[see also,][]{cohn1978,Pau2011,Stone2017,PanZhen2021,Broggi2022}.
The accretion effect will be considered if we introduce the accretion disk in the active stage of a galaxy, where the mass growth is determined by the accretion rate and the accretion time. We assume that the total relative abundance of sBH $\varphi=0.001$ and the density power-law index of $\gamma=1.5$ \citep{Tremaine1994,Binney2008}. The influence radius within which the central MBH dominates the gravitational field is defined from the velocity dispersion of the spheroid of the galaxy, which follows the $M_\bullet-\sigma_*$ relation in equation~\eqref{M_sigma_relation}.

\subsection{Quiet stage}
\begin{figure}
\centering
\includegraphics[width=\columnwidth]{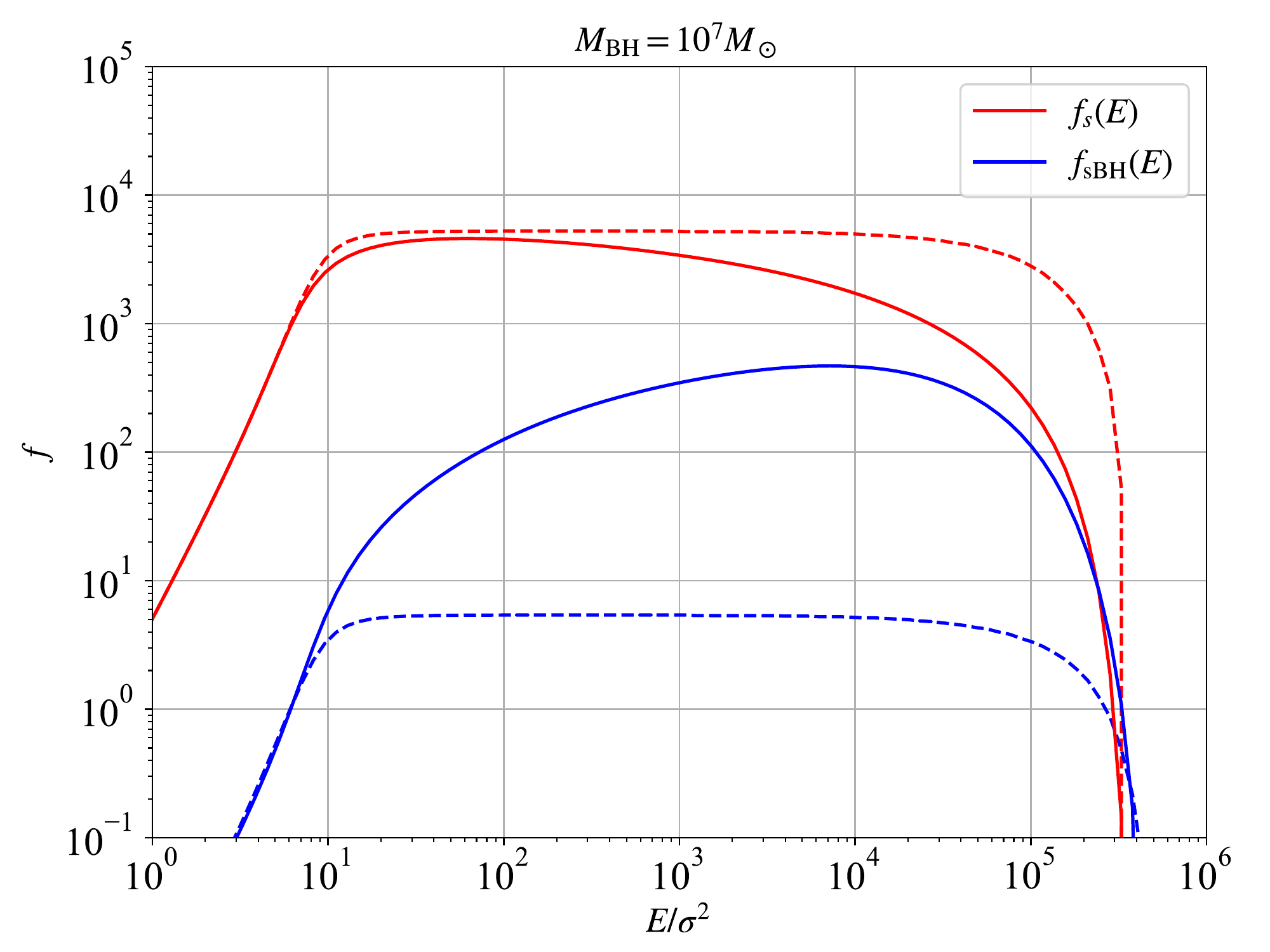}
\caption{Evolution of stars and sBHs during the quiet stage. We plot the marginal distribution function $f_{s/\rm sBH}(E)=\int {\rm d}\,m\, {\rm d}\,\mathscr{R}\,f_{s/\rm sBH}(E,\mathscr{R},m)$ of the stars and sBHs in the unit of ${\rm pc^{-3}}\sigma_*^{-3}$. The blue/red dashed and solid lines represent the initial marginal sBH/star distribution and that after 5\,Gyrs evolution, respectively. The mass of the central MBH is set to be $10^7$\,$M_{\odot}$ in plotting this figure.}
\label{fig:initial_distribution}
\end{figure}

During the quiet stage, the evolution of the distribution function follows the standard way, in which we choose the mass distribution as ${\rm d}N_s/{\rm d}m_s \propto \delta (m_s-M_\odot)$ for stars and ${\rm d}N_{\rm sBH}/{\rm d}m_{\rm sBH} \propto m_{\rm sBH}^{-2.35}$ within $m_{\rm sBH}\in[5,15]\,M_\odot$ for sBHs.
The evolution time is set to be 5\,Gyr for the quiet stage. 
The evolution of the marginal distribution function $f_{s/\rm sBH}(E)=\int {\rm d}\,m \,{\rm d}\,\mathscr{R}\,f_{s/\rm sBH}(E,\mathscr{R},m)$ is plotted in Fig.~\ref{fig:initial_distribution}, while the $f_{s/\rm sBH}(m)=\int {\rm d}\,E {\rm d}\,\mathscr{R}\,f_{s/\rm sBH}(E,\mathscr{R},m)$ is almost unchanged since there is no mass accretion effect during the quiet stage. In Fig.~\ref{fig:initial_distribution}, the density profile of the more massive sBH component for $r<r_h$ is steeper than that of the star component. This is because of the mass segregation effect \citep[][]{Pau2011,Pau2018} which we briefly describe as follows. The total number of stars/sBHs will decrease since they fall into the MBH via the loss cone mechanism. Only in the inner region ($r<r_h$, where $r_h\equiv GM_\bullet/\sigma_\star^2$ is the influence radius of MBH ), the number of stars decreases while the number of sBHs increases. In the inner region, the increase of the sBHs' population is due to the inward migration of the sBHs from the outer region. In the simulation using the Fokker-Planck equation, the heavier sBHs migrate inward faster than the stars, which is known as the mass segregation effect. Thus, the density profile of the more massive sBH component is steeper than that of the star component due to the mass segregation effect.

The results of the quiet stage will be the initial condition for the active stage.

\subsection{Active stage}
For the galaxy entering the active stage, an SG-disk with accretion rate $\dot M_\bullet=0.1 \dot M_{\rm Edd}$ and $\alpha=0.1$ are assumed, the accretion disk will affect both the dynamics of the stars/sBHs and the mass distributions. The features of the evolution during the active stage can be summarised as follows.

\begin{figure}
\centering
\includegraphics[width=\columnwidth]{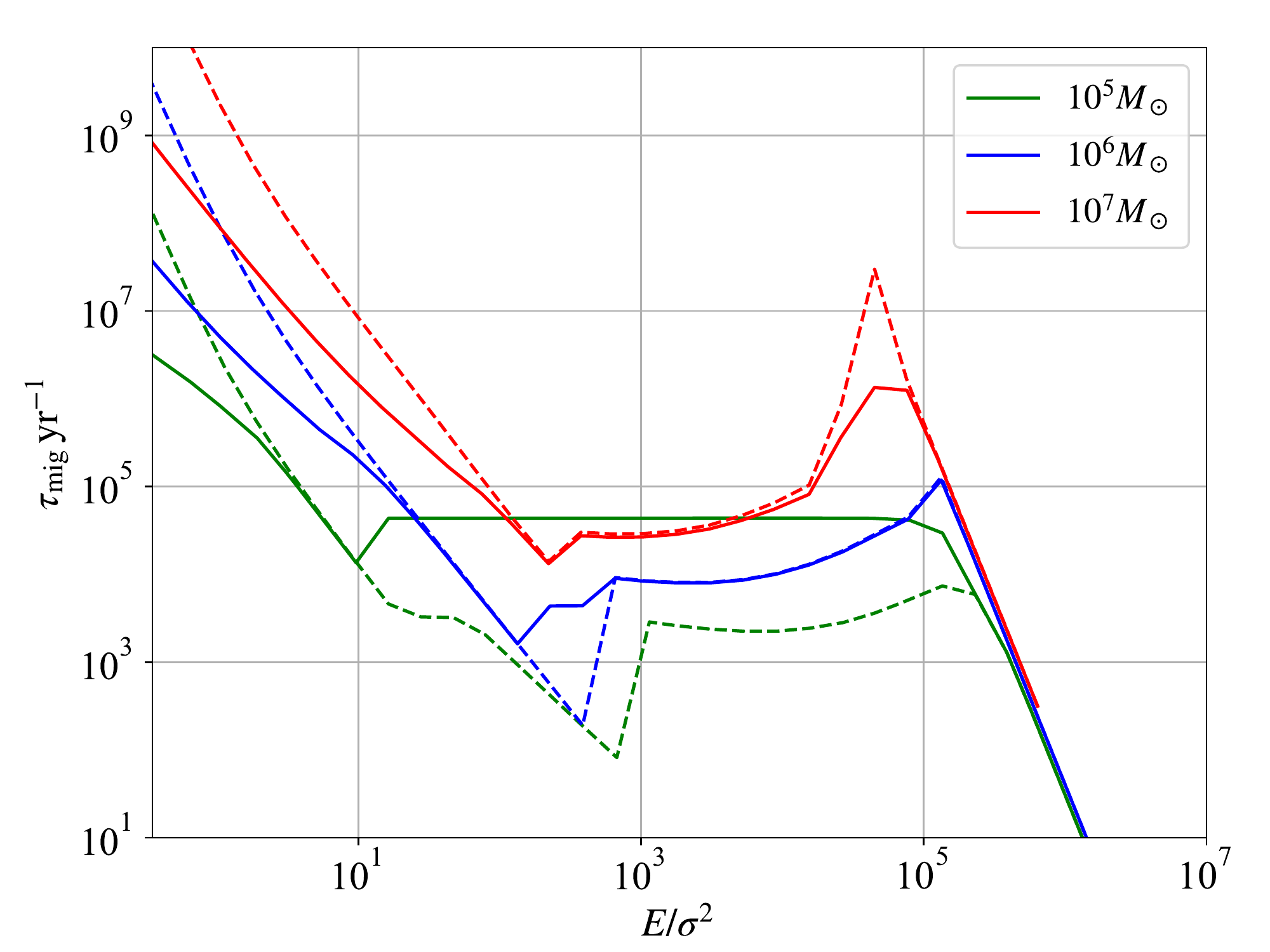}
\caption{Migration time scale $\tau_{\rm mig}(E/\sigma_*^2)$. The dashed lines are the migration time scale for the stars/sBHs on the disk with different central BH masses, while the solid lines are the migration time scale for the stars/sBHs in the nuclear star cluster. The mass of the star and sBH here is chosen to be $10M_\odot$ as fiducial.}
\label{fig:migration_rate}
\end{figure}
\begin{figure*}
\centering
\subfigure[]{
\includegraphics[width=\columnwidth]{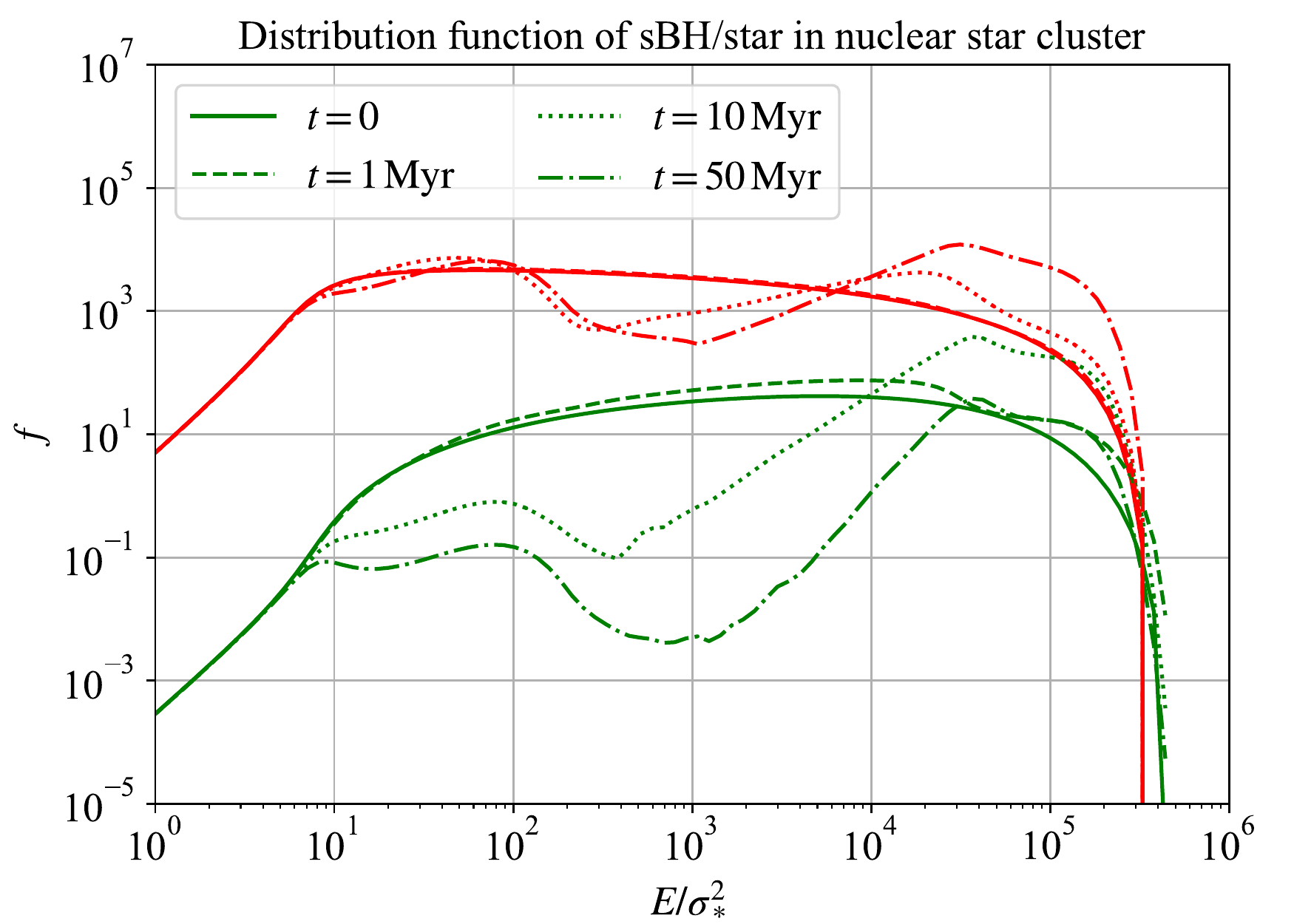}
\label{fig:f_distribution_a}
}
\subfigure[]{
\includegraphics[width=\columnwidth]{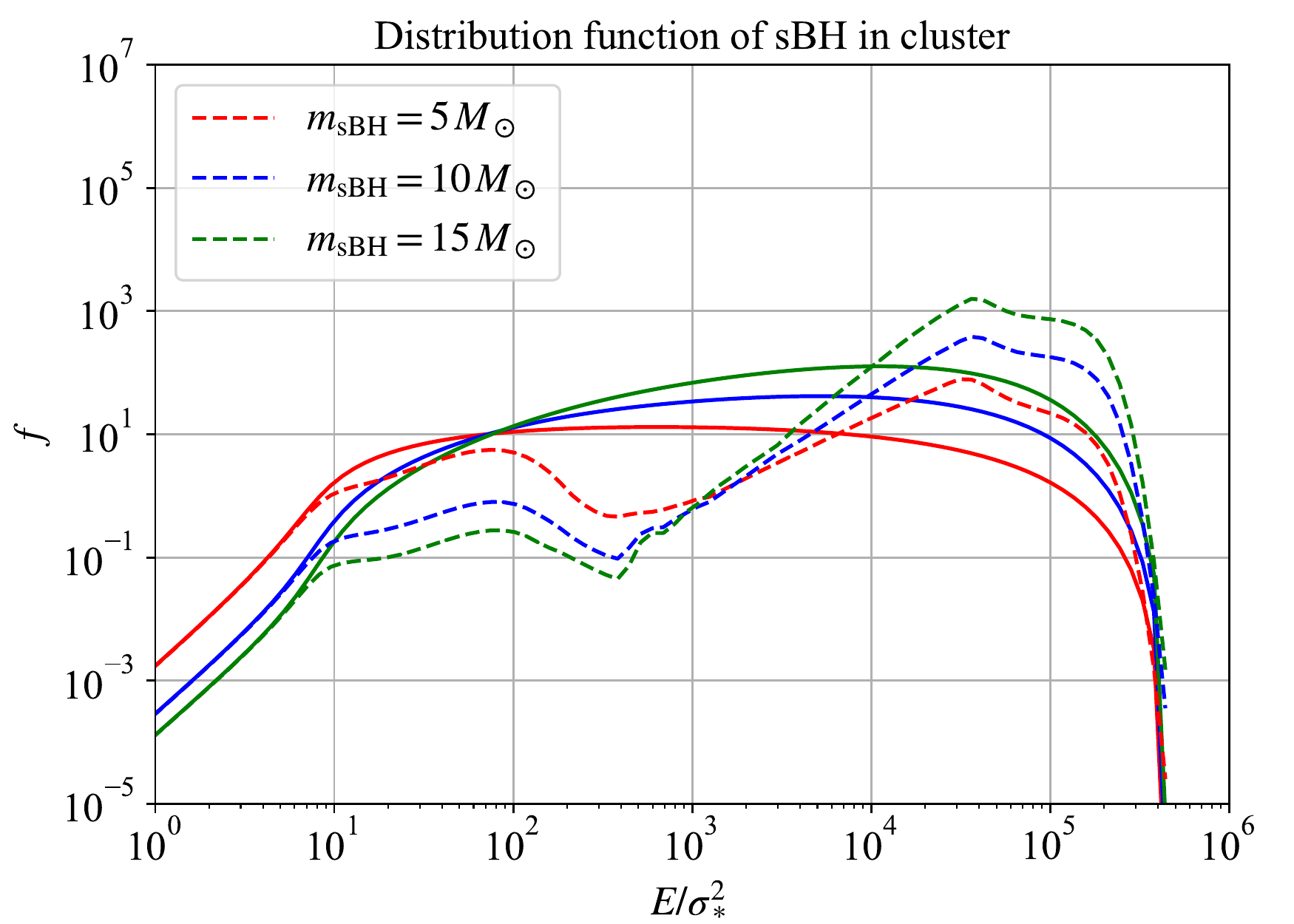}
\label{fig:f_distribution_b}
}
\subfigure[]{
\includegraphics[width=\columnwidth]{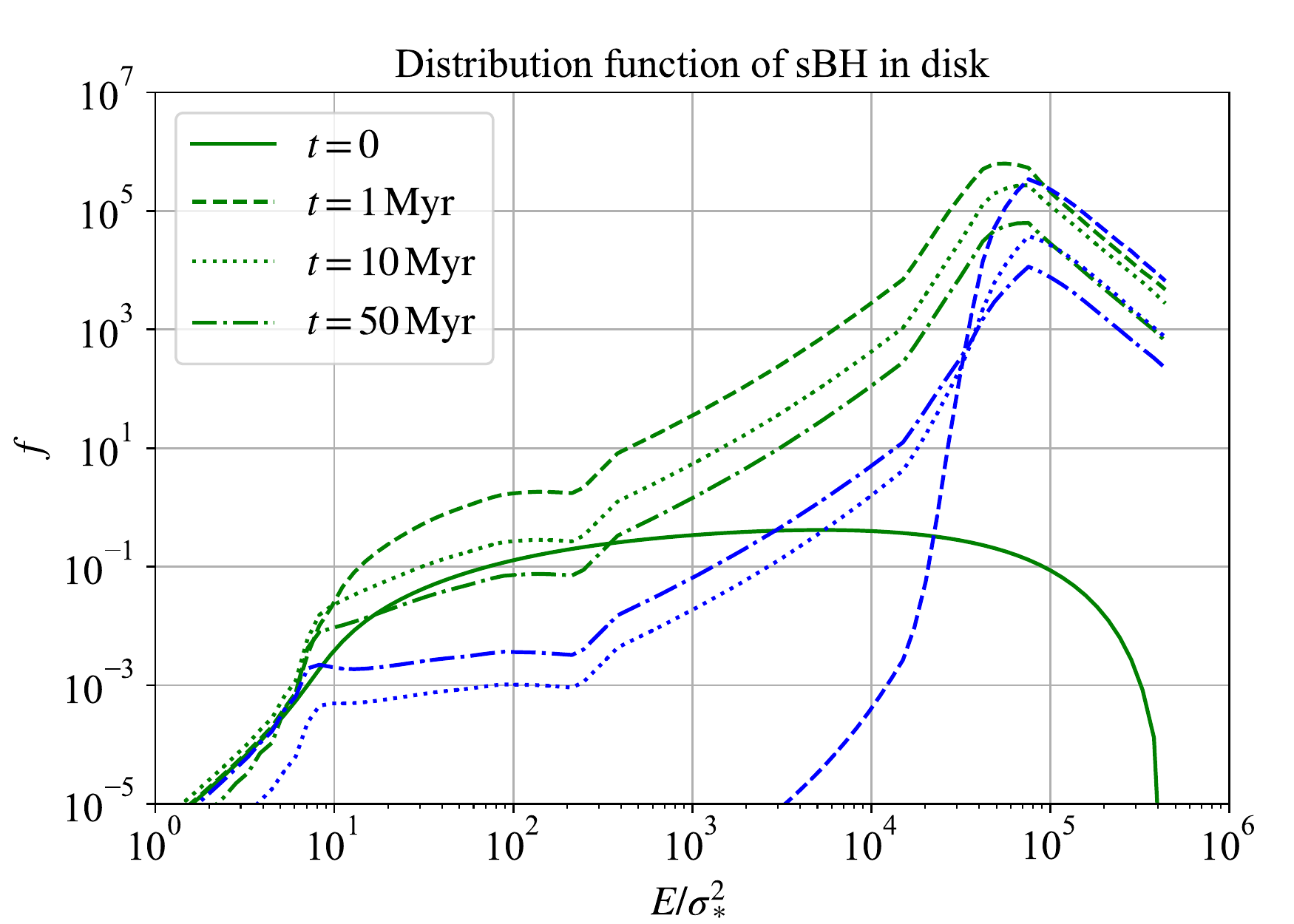}
\label{fig:f_distribution_c}
}
\subfigure[]{
\includegraphics[width=\columnwidth]{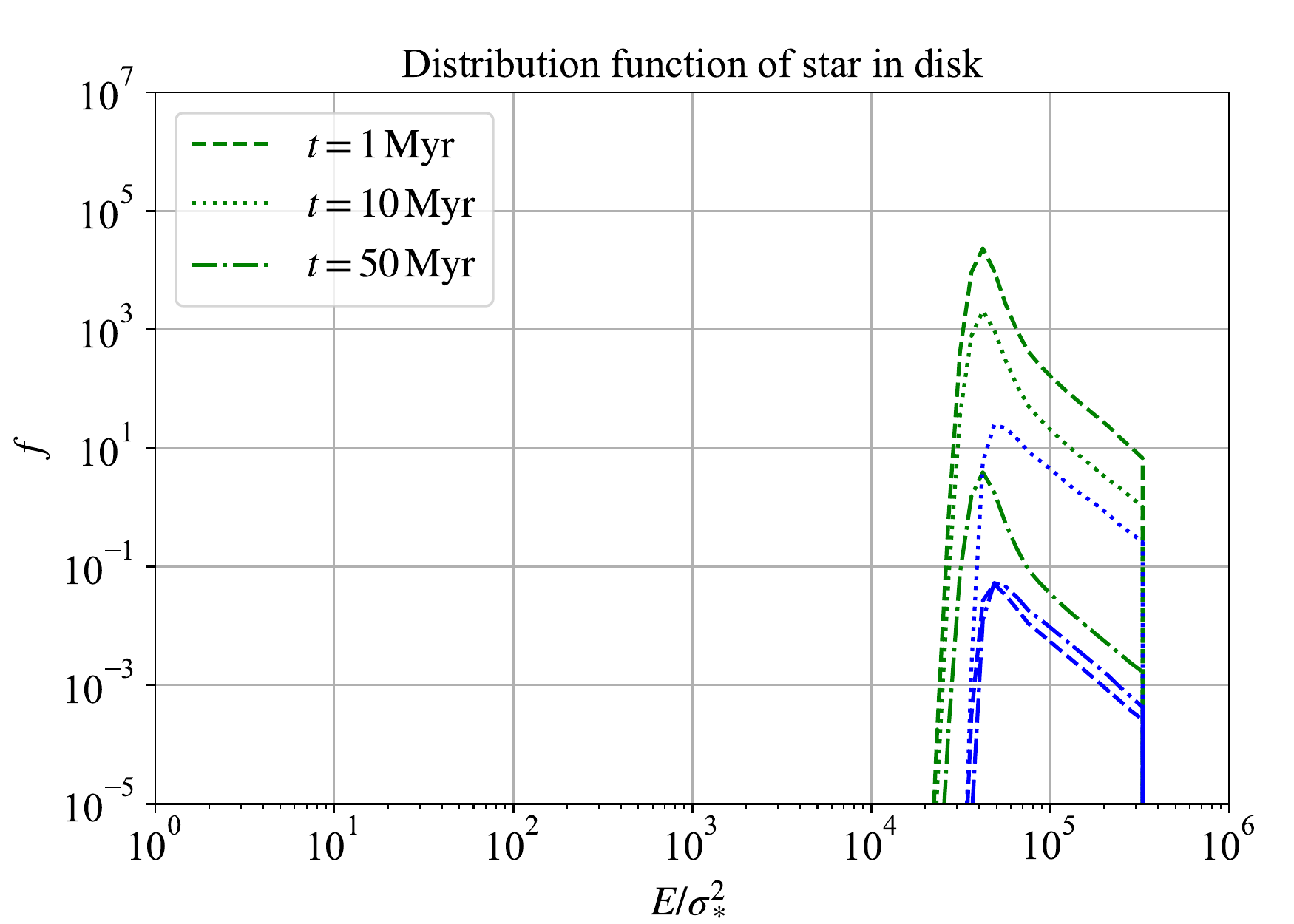}
\label{fig:f_distribution_d}
}
%\includegraphics[width=0.25]{f_distribution_cluster1.pdf}
%\gridline{\fig{f_distribution_cluster1.pdf}{0.5\textwidth}{(a)}
%          \fig{f_distribution_cluster2.pdf}{0.5\textwidth}{(b)}
%          }
%\gridline{\fig{f_distribution_disk1.pdf}{0.5\textwidth}{(c)}
%          \fig{f_distribution_disk2.pdf}{0.5\textwidth}{(d)}
%          }
\caption{The evolution of the distribution function of the stars/sBHs. \quad \emph{Left upper panel}: the evolution of the distribution function for stars/sBHs in nuclear star cluster. The red and green lines represent the distribution function $f_s(t,E)=\int f_s(t,E,m_s)\delta (m_s-M_\odot) \,{\rm d}m_s$ in unit of $\rm pc^{-3}\sigma_*^{-3}$ for stars and $f_{\rm sBH}(t,E,m_{\rm sBH}=10\,M_\odot)$ in the unit of $\rm pc^{-3}\sigma_*^{-3} M_\odot^{-1}$ for sBHs, respectively. \quad \emph{Right upper panel}: the evolution of the sBH's distribution function in nuclear star cluster at $t=10\,\rm Myr$ for different masses of sBHs, where the solid and dashed lines represent the initial distribution and that at $t=10\,\rm Myr$ in the unit of $\rm pc^{-3}\sigma_*^{-3} M_\odot^{-1}$, respectively. \quad \emph{Left lower panel}: the evolution of the sBH's distribution function in AGN disk, where the green and blue lines represent the distribution $f_{\rm sBH}(t,E,m_{\rm sBH})$ in the unit of $\rm pc^{-3}\sigma_*^{-3} M_\odot^{-1}$ with $m_{\rm sBH}=10\,M_\odot$ and $m_{\rm sBH}=20\,M_\odot$, respectively. \quad \emph{Right lower panel}: the evolution of the star's distribution function in AGN disk, where the green and blue lines represent the distribution $f_{\rm sBH}(t,m_s)$ in a unit of $\rm pc^{-3}\sigma_*^{-3} M_\odot^{-1}$ with $m_{s}=50\,M_\odot$ and $m_s=150\,M_\odot$, respectively.  
%The black dotted lines in the above four figures show the boundary of EMRI in a circular orbit contributing to the LISA band.
}

\label{fig:evolution_distribution_function}
\end{figure*}

%For the stars/sBHs in the galaxy cluster, their distribution functions are affected by the radius/energy-dependent migration rate in Fig.~\ref{fig:migration_rate}, the capture of the stars/sBHs by the accretion disk of the centre MBH, and the evolution time. 
The radius/energy-dependent migration timescale shown in Fig.~\ref{fig:migration_rate} has a key effect on the evolution of the distribution functions. The migration rate $\tau_{\rm mig}$ of the cluster stars/sBHs in the outer region ($E/\sigma_*^2<10$) is significantly lower than that in the inner region of the galaxy centre ($E/\sigma_*^2>10^4$). This means that the distribution of the stars/sBHs will not deviate significantly from the initial distribution function in the outer region, while the distribution changes significantly in the inner region due to the strong migration effect. The migration timescale become much shorter at $E/\sigma_*^2\sim 10^3$ (see Fig.~\ref{fig:migration_rate}), which also means a higher capture rate\,(for a fixed $\mu$) since it is proportional to $\mu/\tau^{\rm d}_{\rm mig}$. As shown in Fig.~\ref{fig:f_distribution_a}, we take distribution function of $m_{\rm sBH}=10\,M_\odot$ with $\mu_s=0.1$ and $M_\bullet=10^7\,M_\odot$ as an example. The stars/sBHs distribution functions in the cluster will have a dip within the energy range $10<E/\sigma_*^2<10^4$ due to the capture by the disk, while a peak structure around $10^4<E/\sigma_*^2<10^5$ is due to the traffic effect induced by the longer migration timescale.  The distribution function of sBH in the cluster decreases with time evolution due to the capture by disk, which means that the cluster will provide fewer and fewer sBHs. Moreover, since the migration and capture rate is proportional to the sBH mass, the heavier sBH will have a stronger migration rate and a stronger capture rate, leading to a steeper distribution function at higher energy and a shallower distribution function at fast capture region (see Fig.~\ref{fig:f_distribution_b}).

\begin{figure*}
\centering
\subfigure[]{
\includegraphics[width=\columnwidth]{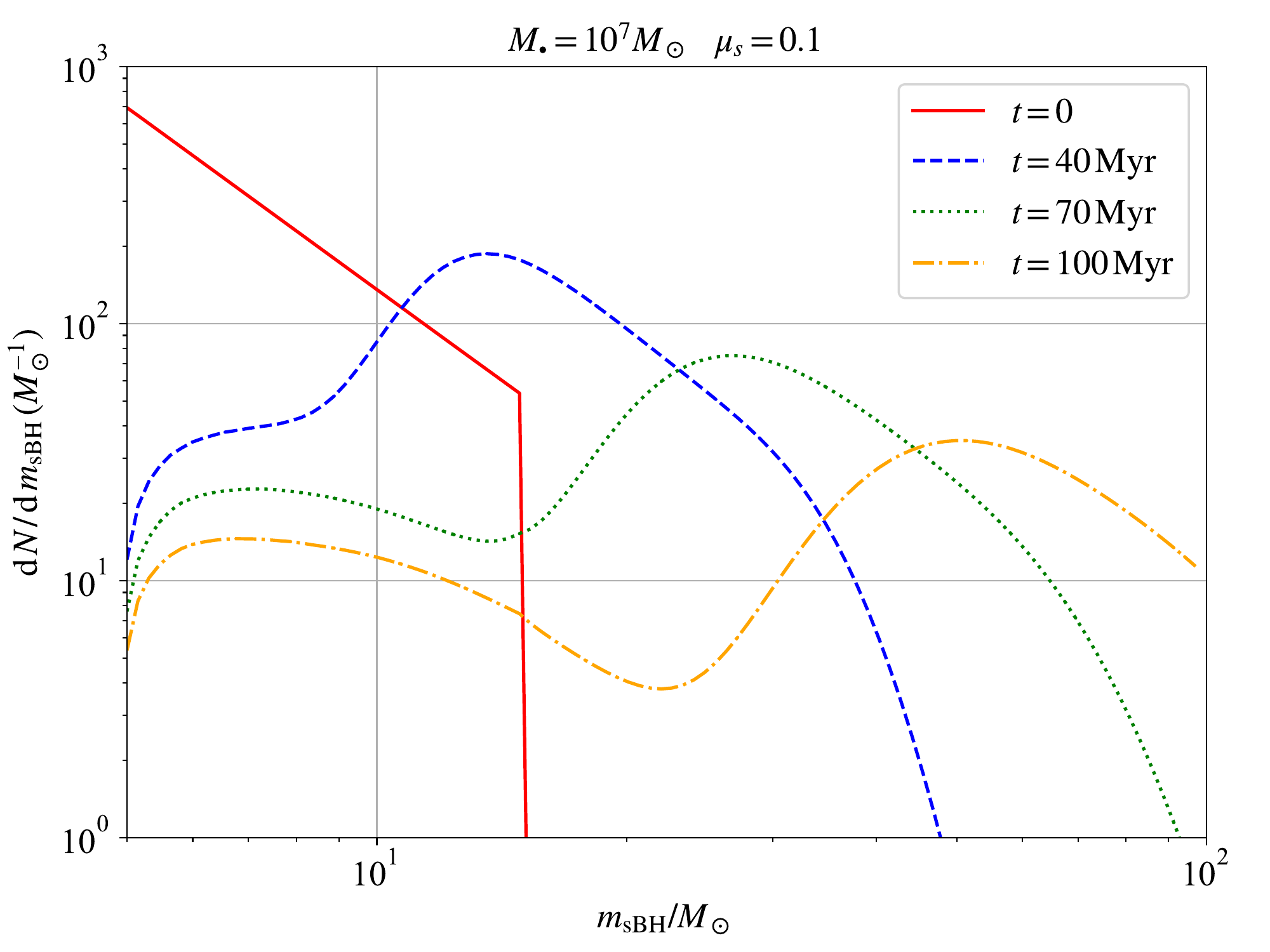}
\label{fig:mass_distribution_a}
}
\subfigure[]{
\includegraphics[width=\columnwidth]{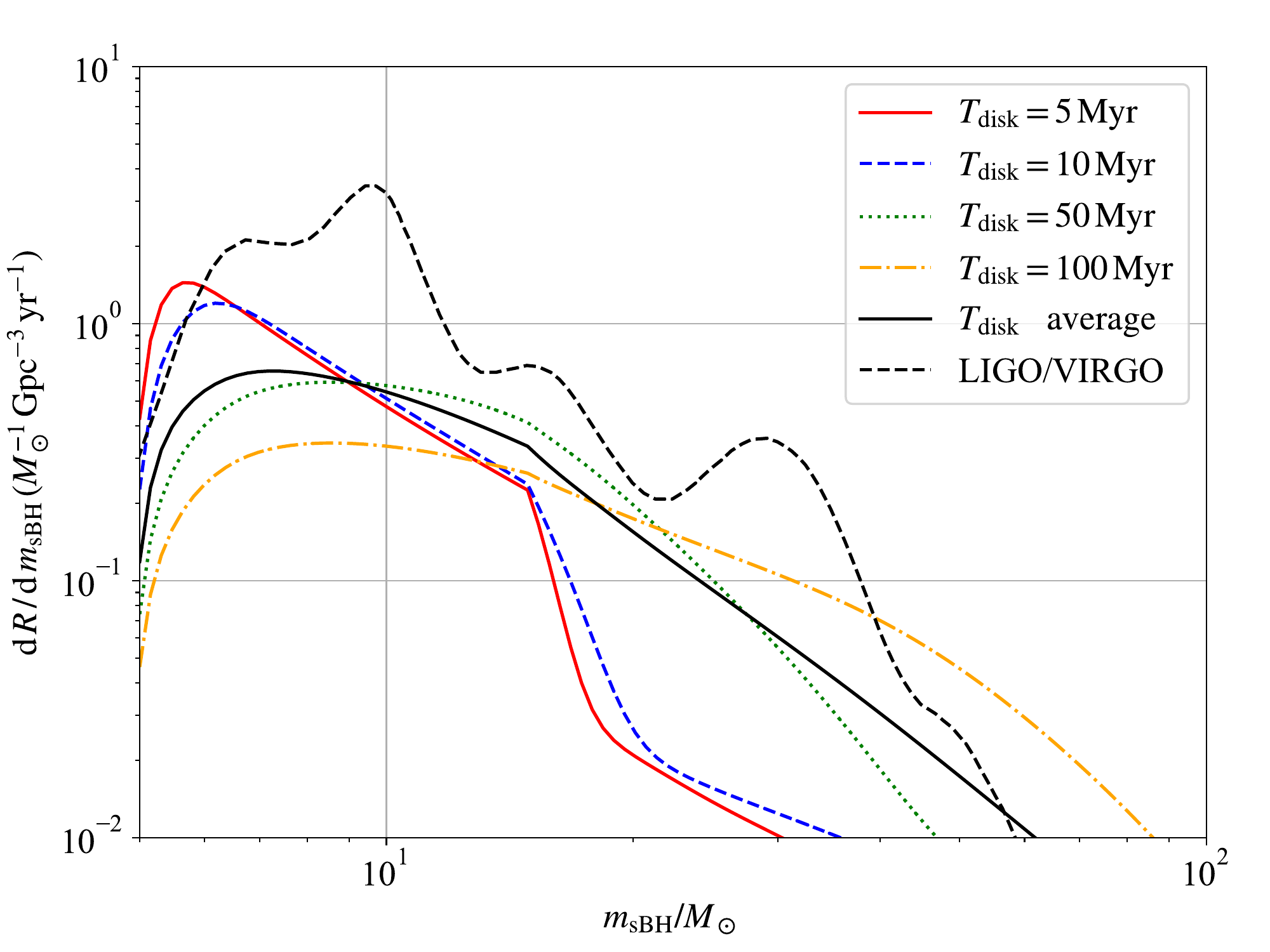}
\label{fig:mass_distribution_b}
}
\caption{The mass distribution of sBH in the disk. \quad \emph{Left panel}: an example of the mass-distribution evolution for different timescales for a MBH with $M_\bullet=10^7M_\odot$ and capture rate $\mu_s=0.1$, where $\dot{M}=\dot{M}_{\rm Edd}$ and $\dot{M}=20\,\dot{M}_{\rm Edd}$ are adopted for sBHs in the outer gravitationally unstable region and inner standard disk region, respectively. \quad \emph{Right panel}: the number density distribution of sBH binary merger event. The black solid line represents the average distribution over different disk lifetimes within $T_{\rm disk}\in [1,100]\,\rm Myr$. For comparison, the black dashed line represents the mass distribution reconstructed from the merging BBHs as observed by LIGO/Virgo.}  
\label{fig:mass_distribution}
\end{figure*}

In Fig.~\ref{fig:f_distribution_c}, we show the distribution function of $m_{\rm sBH}=10\,M_\odot$ and $20\,M_\odot$ in the disk during different evolution stages. The peaks of these distribution functions gradually shift to the high energy region until it reaches local equilibrium, i.e., the capture rate from the cluster is equal to the inward migration rate. The peak of the distribution function around $E/\sigma_*^2\approx 5\times 10^4$ in the equilibrium state is due to the traffic effect and it is consistent with the peak in the figure of migration timescale\,Fig.~\ref{fig:migration_rate}. Due to the accretion, the distribution function of lighter sBHs gradually decreases while the distribution function of heavier sBHs gradually increases. It is worth mentioning that, evolving to the equilibrium stage, the total number of sBHs in the disk decreases with time since some of them plunge into the centre MBH and the capture rate from the cluster also decreases with time. In this way, the wet EMRI rate will also decay over time, which is consistent with the result in \citet{PanZhen2022}. Similarly, the distribution function of stars with $m_s=50\,M_\odot$ and $150\,M_\odot$ is shown in Fig.~\ref{fig:f_distribution_d}. In the unstable region of disk, the stars can not grow to $50M_\odot$ due to their low accretion rate ($\dot{m}_s=\dot{M}_{\rm Edd}$). In the stable region of disk, 
only the stars around the peak of migration timescale $E/\sigma_*^2\approx 5\times 10^4$ have enough time to become so massive.

The above information on the evolution of distribution functions of the stars/sBHs could be useful to understand the black hole distribution in the AGN disk and also the gravitational waves emitted from the corresponding EMRI events, as we shall discuss in the next sections.

\section{Mass distribution of sBHs in disk}\label{sec:5}
% Relation to the Black holes observed by LIGO

Up to now, more than 90 stellar-mass BBH merging events have been detected by Advanced LIGO \citep{LIGO2015} and Advanced Virgo \citep{Virgo2015}. However, the origin of these sBH binaries is still unclear. The accretion disk provides a unique environment for BBHs to merge \citep[see][for a review]{Mandel2022}. In such a gas-rich environment, the disk-binary interaction will accelerate the binary merge \citep[e.g.,][]{Grobner2020}. In addition, the migrating object will be trapped in the location where $\dot J_{\rm mig,I}$ changes its sign from negative to positive in the decreasing $r$ direction, which is called migration trap \citep[e.g.,][]{Lyra2010,McKernan2012,McKernan2014,Bellovary2016ApJ,Secunda2019}. These migration traps are favorable places for BBH formation \citep[e.g.,][]{Secunda2019,Peng2021}. %BAD ENGLISH AND BAD LOGIC!
 Various works\,\citep[e.g.,][]{Stone2017,Yang2019apj,Yang2019prl,Grobner2020,Gerosa2021} suggested that the AGN-assisted BBH merger rate is about $(0.02-60) \,\rm Gpc^{-3} yr^{-1}$, which could be responsible for $10\% - 50\%$ of gravitational-wave detections. Meanwhile, of the BBH mergers that have been detected, many sBHs are bigger than $30M_\odot$, which are much larger than the remnants of traditional stellar evolution ($\sim 10M_\odot$). These massive sBHs are usually regarded as the remnants of the hierarchical sBH mergers \citep[see][]{Kohler2017,Yang2019prl}. Here, we expect that accretion could offer another channel to form massive sBHs detected by LIGO/Virgo. 

Moreover, future upgraded ground-based GW detectors will detect a large amount of binary sBH merger events. A sub-species of these binary merger events may have EM counterparts if they happen in the AGN disk. It should be noted that it is difficult to find out this kind of EM counterpart due to the bright background emission from the AGN disk. Up to now, ten possible candidate EM counterparts to BBH mergers are reported in \citet{Graham2020,Graham2022}.   Therefore it is worth investigating the effect of the accretion of the sBH in the disk on the mass distribution of these BBH merger events. On the other hand, multi-messenger observations in the future on these sources will be beneficial to the understanding of the galaxy centre environment.

In Fig.~\ref{fig:mass_distribution_a}, we show a typical mass distribution during the active stage where the accretion rate $\dot{m}_{\rm sBH}=20\,\dot{M}_{\rm Edd}$ in the inner region of disk and $\dot{m}_{\rm sBH}=\dot{M}_{\rm Edd}$ in the outer region. The peak gradually shifts to the more massive region over time due to mass growth by accretion. Some sBHs can even grow up to $100M_\odot$, which is determined by their accretion rate and the duration of the migration process in the disk.

Detection of a BBH merger event can be understood as sampling over many host galaxies with different parameters, such as the evolution stage of the galaxy, the galaxy's centre MBH mass, disk lifetime (duty cycle) $T_{\rm disk}$, etc. To obtain the number density distribution ${\rm d}\,n/{\rm d}m_{\rm sBH}$ of sBH in the AGN disk relevant to the future GW detection, we need to average our number density of the sBH over the centre MBH mass $M_{\bullet}\in [10^5-10^{10}]\,M_\odot$, evolution time $t\in[0, T_{\rm disk}]$ and the disk lifetime $T_{\rm disk}\in[T_{\rm disk}^{\rm min},T_{\rm disk}^{\rm max}]$. Firstly, we can compute the number density per unit sBH mass with a fixed duty cycle defined as:
\be
\frac{{\rm d}n}{{\rm d}m_{\rm sBH}}=\int_0^{T_{\rm disk}} \frac{{\rm d}\,t}{T_{\rm disk}} \int^{10^{10}\,M_\odot}_{10^5\,M_\odot} {\rm d}\,{\rm log} M_\bullet\, \frac{{\rm d}\,N(M_\bullet,t)}{{\rm d}\,m_{\rm sBH}}\phi_{\rm bh}\,f_{\rm wet},
\ee
where $\phi_{\rm bh}$ is the local MBH mass function \citep[see Figure\,5 in][]{Barausse2012} and the fraction of AGN $f_{\rm wet}$ is assumed as $1\%$. Furthermore, we can estimate the distribution of BBH mergers over mass in AGN disk as \citep[][]{Grobner2020}
\be\label{eq:binary_converter}
\frac{{\rm d}\,R}{{\rm d}\,m_{\rm sBH}}=\frac{{\rm d}\,n}{{\rm d}\,m_{\rm sBH}}\frac{f_b}{\tau_{\rm med}},
\ee
where we set $f_b=0.1$ as a typical fraction of such sBHs residing in the binaries \citep[$f_b\sim 0.01-0.2$ in][]{McKernan2018} and $\tau_{\rm med}=7\times 10^6\,\rm yr$ as a typical median BBHs merger timescale in the disk \citep[$\tau_{\rm med}\sim 10^6-10^8\,\rm yr$, which depends on disk parameters and sBH binaries' initial orbital parameters, see][]{Grobner2020}. We show the results in red solid, blue dashed, green dotted and orange dash-dotted lines for $T_{\rm disk}=\{5,10,50,100\}\,\rm Myr$ in Fig.~\ref{fig:mass_distribution_b}.
Moreover, we show the ensemble average in black solid line over the AGN disk lifetime $T_{\rm disk}$, i.e.,
\be
\frac{{\rm d}\bar{R}}{{\rm d}m_{\rm sBH}}=\frac{1}{T_{\rm disk}^{\rm max}-T_{\rm disk}^{\rm min}}\int_{T_{\rm disk}^{\rm min}}^{T_{\rm disk}^{\rm max}} {\rm d} T_{\rm disk}\frac{{\rm d}R}{{\rm d}m_{\rm sBH}},
\ee
where $T_{\rm disk}^{\rm min}=1\,\rm Myr$ and $T_{\rm disk}^{\rm max}=100\,\rm Myr$ \citep[e.g.,][]{Shulevski2015,Turner2018}. For comparison, we also show the BBH merger event distribution over mass reconstructed from the merging BBHs observed by LIGO/Virgo \citep{Tiwari2021} in the black dashed line. Our results show that the AGN-assisted BBH merger rate in the disk could be responsible for $10\% - 30\%$ of gravitational-wave detections. The peak of binary sBH mass distribution observed by the ground based gravitational wave detector is at about $9\,M_\odot$ \citep[see][]{Tiwari2021,VanSon2022}. The accretion effect of sBHs in the disk in our model could be one possible reason that contributes to the dearth of low-mass binary sBH by shifting the sBHs population to the more massive region.

%However, the formation mechanism of the detected BBHs is currently not unclear.  }
 
\section{EMRI event rate}\label{sec:6}
The above simulation results of the distribution function evolution can be used to study the influence of the accretion on the EMRI event rate, which will be related to the stochastic GWB contributed by EMRI events in the next section. Following the definition in \citet{PanZhen2021wet}, we name EMRIs in quiet/active stage as dry/wet EMRIs with event rates denoted as $\Gamma_{\rm dry/wet}$, respectively. For the dry EMRI per MBH with sBH mass $m_{\rm sBH}$, the EMRI rate via loss cone mechanism is given by  \citep[e.g.,][]{Hopman2005,Pau2011,PanZhen2021}
\be 
\Gamma_{\rm dry}(m_{\rm sBH})=\int_{E>E_{\rm gw}} \vec{F}\cdot {\rm d}\vec{l}
\ee
where $E_{\rm gw}=GM_\bullet/2r_{\rm gw}$ with $r_{\rm gw}=0.01 r_h$, $\vec{F}=\{F_E,F_R\}$ is the flux of star/sBH along the line element of the loss cone boundary ${\rm d}\vec{l}=\{{\rm d}E,{\rm d}R\}$. It is important to note that the real observation is over an ensemble of EMRI sources with different evolution times. Therefore we need to have a time/ensemble averaged event rate, hence the total mean EMRI rate overall sBH masses averaged within time $T$ is defined as:
\be\label{eq:averaged_EMRI}
\bar\Gamma=\int {\rm d}m_{\rm sBH}  \int_{0}^{T} \frac{{\rm d}t}{T}\, \Gamma(m_{\rm sBH},t).
\ee

For simplicity, we take the fitting formula obtained by\,\citet{Broggi2022} for the mean dry EMRI rate, where they assumed $m_{\rm sBH}=10\,M_\odot$ and suggested that the best-fit to the mean dry EMRI rate within a Hubble time $T_{H}\sim 10\,\rm Gyr$:
\be \label{eq:dry_EMRI}
\bar\Gamma_{\rm dry}=130\left(\frac{M_\bullet}{4\times 10^6M_\odot}\right)^{1.03}\, \rm Gyr^{-1}.
\ee
Note that the dry EMRI rate reproduced in our calculation shows that the above $\bar\Gamma_{\rm dry}$ is only slightly affected by extending the monochromatic distribution\,($m_{\rm sBH}=10M_\odot$) to the power-law distribution\,($dN/dm_{\rm sBH}\propto m_{\rm sBH}^{-2.35}$ with $5\,M_{\odot}<m_{\rm sBH}<15\,M_\odot$). Therefore, we make use of the above equation for the dry EMRI rate in our work. In the active stage of the galaxy, the wet EMRI rate is proportional to the distribution function/number density of sBHs around the $E_{\rm max}$ boundary. Furthermore, the sBHs' number density of the disk component is several orders of magnitude higher than that of the cluster component at $E_{\rm max}$ boundary (see Fig.~\ref{fig:f_distribution_a} and Fig.~\ref{fig:f_distribution_c} ). In this case, the wet EMRI rate will be dominated by the sBHs in the disk, which can be given by 
\be
\Gamma_{\rm wet}(m_{\rm sBH})=\left.-\mathcal{C}D_E^{d}\,g_{\rm sBH}\right|_{E=E_{\rm max}}.
\ee
It should be noted that the wet EMRI rate is determined by the $E-$direction flux at the $E_{\rm max}$ boundary while $R-$direction flux is no longer important due to the strong orbit circularisation in the active stage. To obtain the averaged wet EMRI rate via equation~\eqref{eq:averaged_EMRI}, the time average should be over the disk lifetime thereby $T=T_{\rm disk}$.% We also summarise over 20\,sBH mass bins in the $\log{m_{\rm sBH}}$-units within $m_{\rm sBH}=5-100\,M_\odot$ for calculating $\bar\Gamma_{\rm wet}(m_{\rm sBH})$.

\begin{figure}%\label{fig:EMRI_time}
\centering
\includegraphics[width=0.49\textwidth]{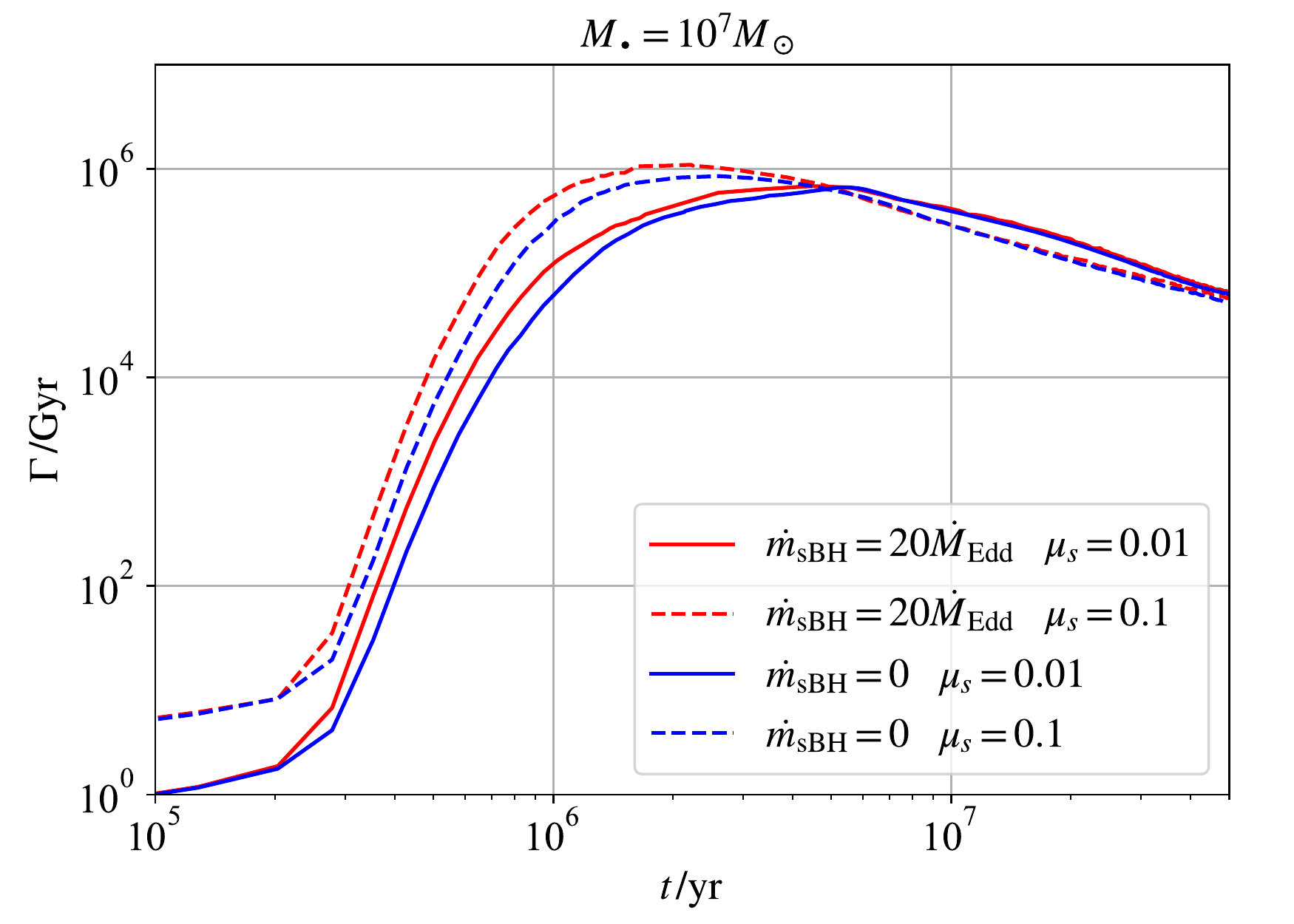}
\caption{The time dependence of wet EMRI rate. The solid and dashed lines represent the capture rate of $\mu_s=0.01$ and $\mu_s=0.1$ respectively. The blue and red lines show the accretion rate of $\dot m_{\rm sBH}=0$ and $\dot m_{\rm sBH}=20\,\dot{M}_{\rm Edd}$, respectively. }
\label{fig:EMRI_time}
\end{figure}

\begin{figure}
\centering
\includegraphics[width=0.49\textwidth]{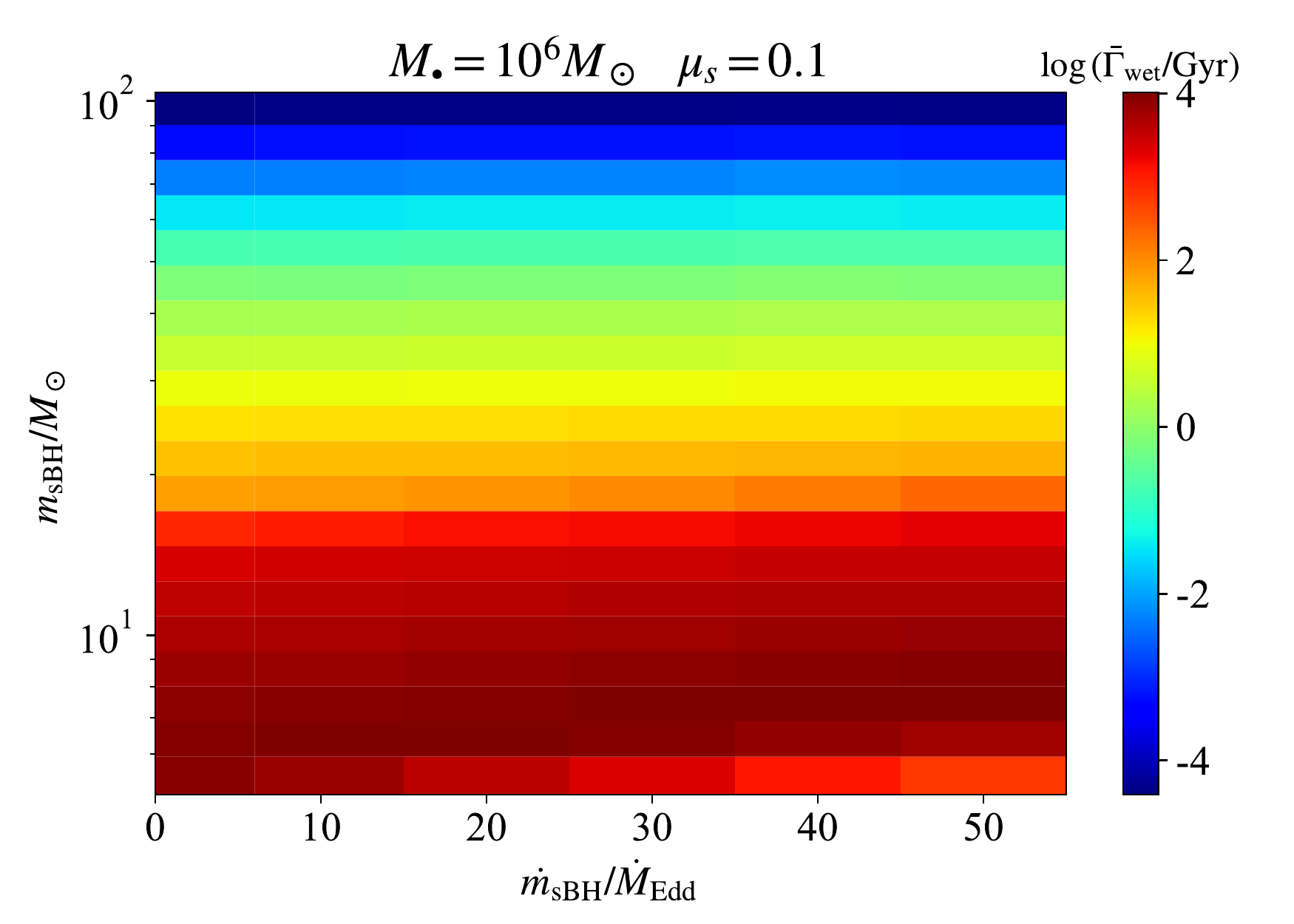}
\includegraphics[width=0.49\textwidth]{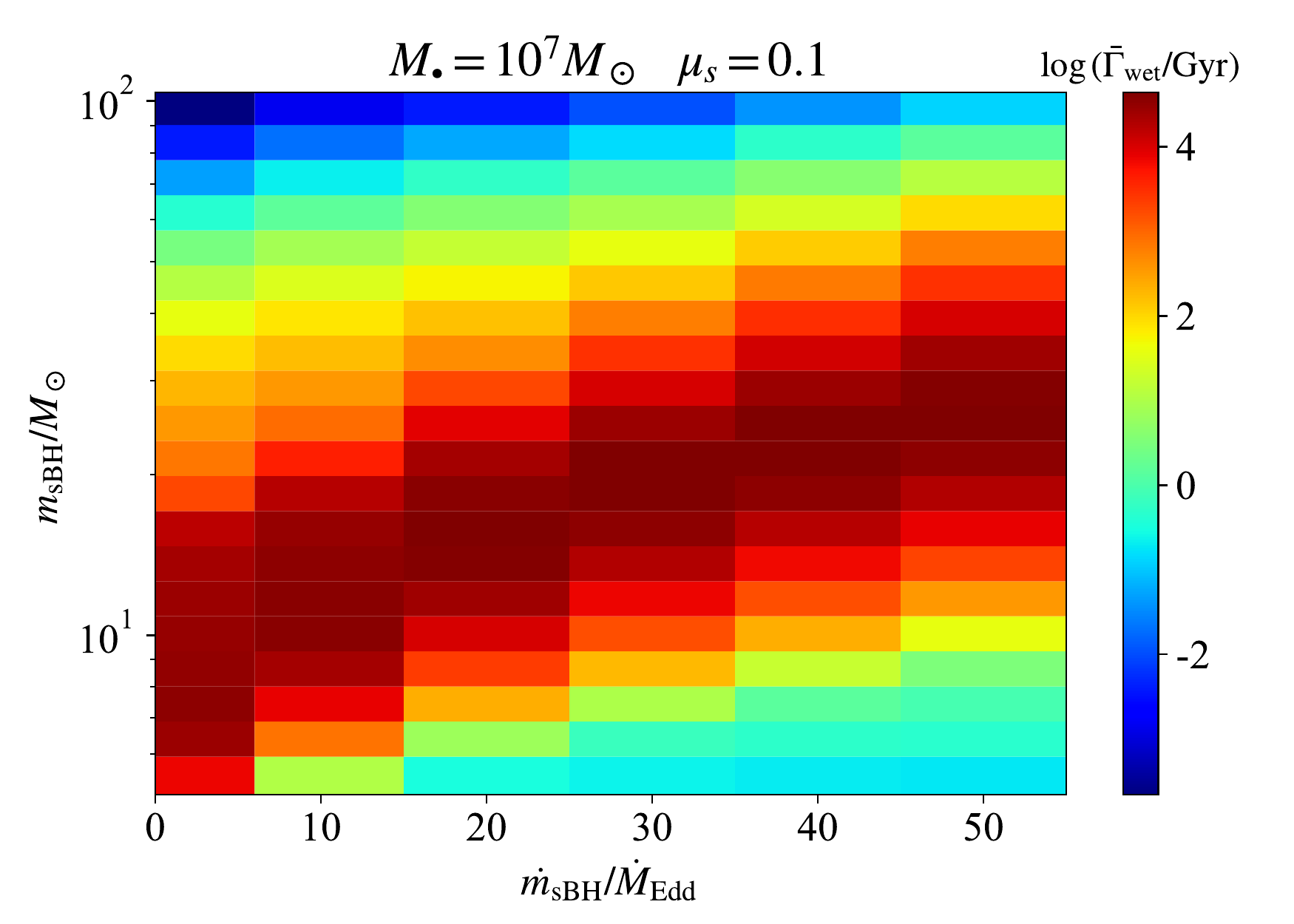}
\caption{The time-averaged wet EMRI rate within $50 \rm Myr$ for different masses of the sBHs that influenced by different accretion rates $\dot{m}_{\rm sBH}=\{2,10,20,30,40,50\}\dot{M}_{\rm Edd}$. The color bar is the time-averaged wet EMRI rate $\bar \Gamma_{\rm wet}=\int_{m^i_{\rm sBH}}^{m_{\rm sBH}^{i+1}} \bar\Gamma_{\rm wet}(m_{\rm sBH})d m_{\rm sBH}$, where $m_{\rm sBH}^i$ is i-th sBH mass bin of $5-100M_\odot$ in the log-scale. The upper and lower panels represent the cases for $M_\bullet=10^6M_\odot$ and $M_\bullet=10^7M_\odot$ with $\mu_s=0.1$, respectively.  }
\label{fig:Wet_EMRI_Rate}
\end{figure}

We show the time dependence total wet EMRI rate $\bar{\Gamma}_{\rm wet}=\int \bar{\Gamma}_{\rm wet}(m_{\rm sBH})\, dm_{\rm sBH}$ for $\mu_s=\{0.01,\, 0.1\}$ with different sBH accretion rates in Fig.~\ref{fig:EMRI_time}. 
For a fixed $\mu_{s/\rm sBH}$, the accretion has a slight promotion on the EMRI rate at the early and middle stages of the evolution. This is because the accretion increases the mass of the sBH, thereby increasing the migration rate. Meanwhile, during the duty cycle of disk, there are about 70 sBHs formed by collapsing stars, which can also increase less than 10\% of the EMRI rate.
After reaching a peak, the event rate will gradually decrease since more and more sBHs plunged into the MBH. The capture rate has little effect on a long-time averaged\,(mean) EMRI rate ($\bar \Gamma_{\rm wet}$ with $t>10\,\rm Myr$) since the total number of sBHs is the same at different capture rates, capture rate only affects how the $\Gamma$ evolves with time. 
However, the mass accretion of sBH can affect the mass distribution of sBH in the EMRI system. In Fig.~\ref{fig:Wet_EMRI_Rate}, we show how the total mean wet EMRI rate $\bar\Gamma_{\rm wet}$ within $m_{\rm sBH}^i<m_{\rm sBH}<m_{\rm sBH}^{i+1}$  depends on the accretion rate of the sBH for two systems with different MBH masses. Since for low-mass AGN, the migration time of sBH is comparatively short, the accretion effect on sBH does not have enough time to accumulate and hence has little effect on the wet EMRI rate\,(see the upper panel of Fig.~\ref{fig:Wet_EMRI_Rate}). On the other hand, the accretion effect is not negligible in massive galaxies (see the lower panel of Fig.~\ref{fig:Wet_EMRI_Rate}), which will significantly increase the wet EMRI events consisting of heavier sBH and MBH. 

In general, the wet EMRI rate is 2-3 orders of magnitude higher than the dry EMRI rate. For a single EMRI source, the dry and wet EMRI events could be distinguishable by observing the GW waveform due to their different eccentricity. The wet EMRI triggered by star-disk interaction tends to be in a circular orbit while the dry EMRI triggered by muti-body scattering tends to be in an elliptical orbit \citep[][]{Babak2017,Pau2018,Bonetti2020}. The waveform of EMRI in an eccentric orbit will be eccentric GW bursts \citep[e.g.,][]{Loutrel2020b,Loutrel2020a}, while the EMRI in a circular orbit has an approximate sinusoidal GW waveform.

%\R{In Figure\,\ref{fig:Wet_EMRI_Rate}, we show the relationship between wet EMRI Rate for 20 sBH mass bins in $\log-$space within $m_{\rm sBH}=5-100M_\odot$ and the accretion rate of sBH where $\bar\Gamma$ is defined by}
%\be
%\bar\Gamma=\int dm_{\rm sBH}  \int_{0}^{T_{\rm disk}} \frac{dt}{T_{\rm disk}}\, %\Gamma(m_{\rm sBH},t)
%\ee
%\R{where we accept a typical AGN lifetime $T_{\rm disk}=5\times 10^7 \rm yr$. Due to the short migration time scale of low-mass AGN (see Figure\,\ref{fig:Wet_EMRI_Rate}(a)), the accretion process of sBH has little effect on the wet EMRI rate. On the other hand, the accretion effect is not negligible in massive galaxies (see Figure\,\ref{fig:Wet_EMRI_Rate}(b)), which will significantly increase the event rate of massive sBH.}\\

%The time average wet EMRI rate within $5\times 10^7 \rm yr$ for sBH accretion rate $\dot{m}_{\rm sBH}=\{2,10,20,30,40,50\}\dot{M}_{\rm Edd}$. The colorbar is time average wet EMRI rate $\bar \Gamma_{\rm wet}=\int_{m^i_{\rm sBH}}^{m_{\rm sBH}^{i+1}} \bar\Gamma_{\rm wet}(m_{\rm sBH})d m_{\rm sBH}$, where $m_{\rm sBH}^i$ is i-th sBH mass bin of $5-100M_\odot$ in log space. The upper and lower panel show the condition of $M_\bullet=10^6M_\odot$ and $M_\bullet=10^7M_\odot$ with $\mu_s=0.1$, respectively.}

\section{Stochastic gravitational wave background in milli-Hertz band}\label{sec:7}
In this section, we present the stochastic gravitational wave background of wet EMRI. The wet EMRI rate will be 2-3 orders of magnitude higher than the dry EMRI due to star-disk interactions, which means a stronger GWB. Moreover, we also include the effect of sBHs' accretion on the GWB.  
%We chose the LISA sensitivity as a fiducial and 
We follow the calculation procedure in \citet{Bonetti2020}, which is briefly summarised as follows.\\

\begin{figure*}%\label{fig:GWB}
\centering
\includegraphics[width=\columnwidth]{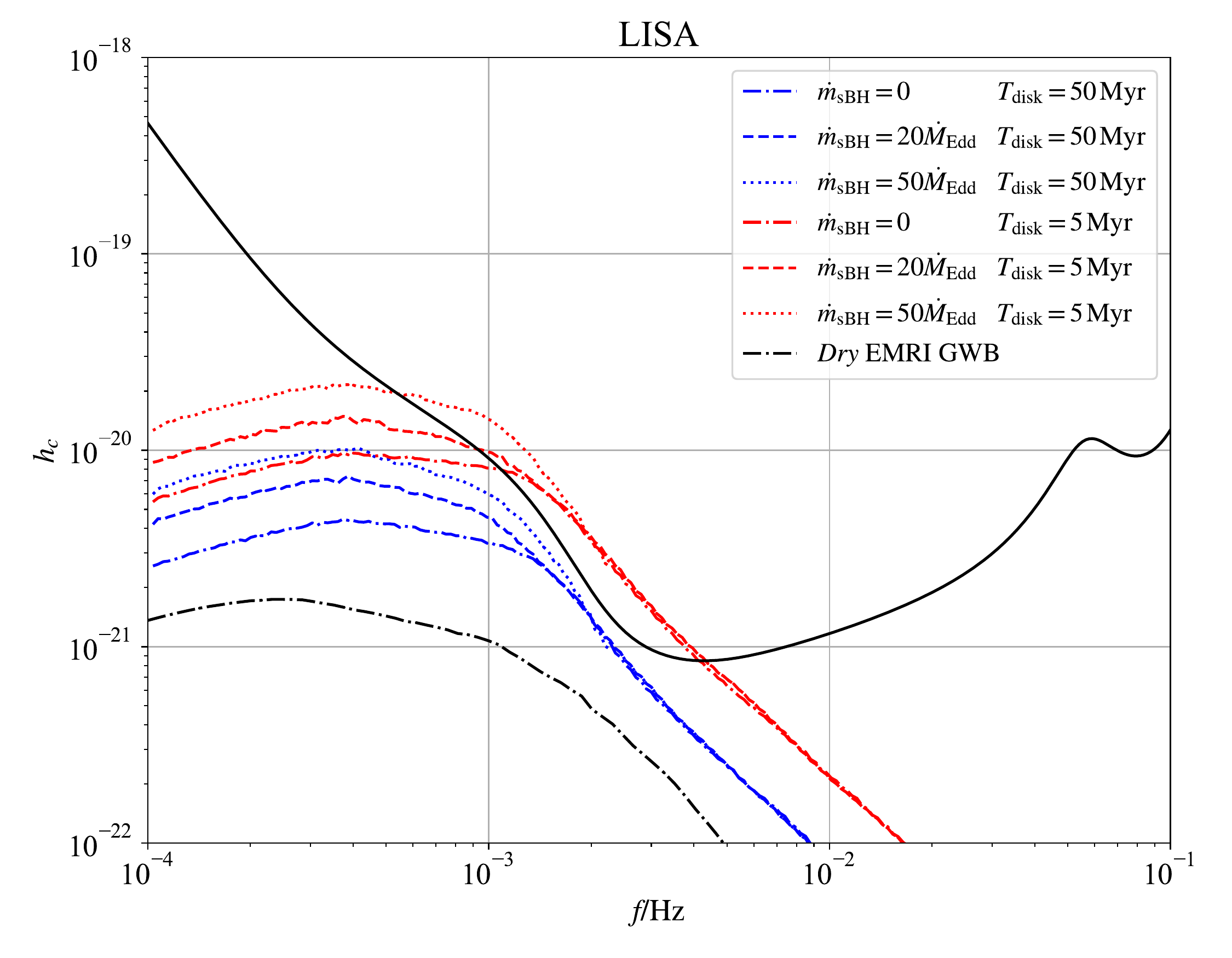}
\includegraphics[width=\columnwidth]{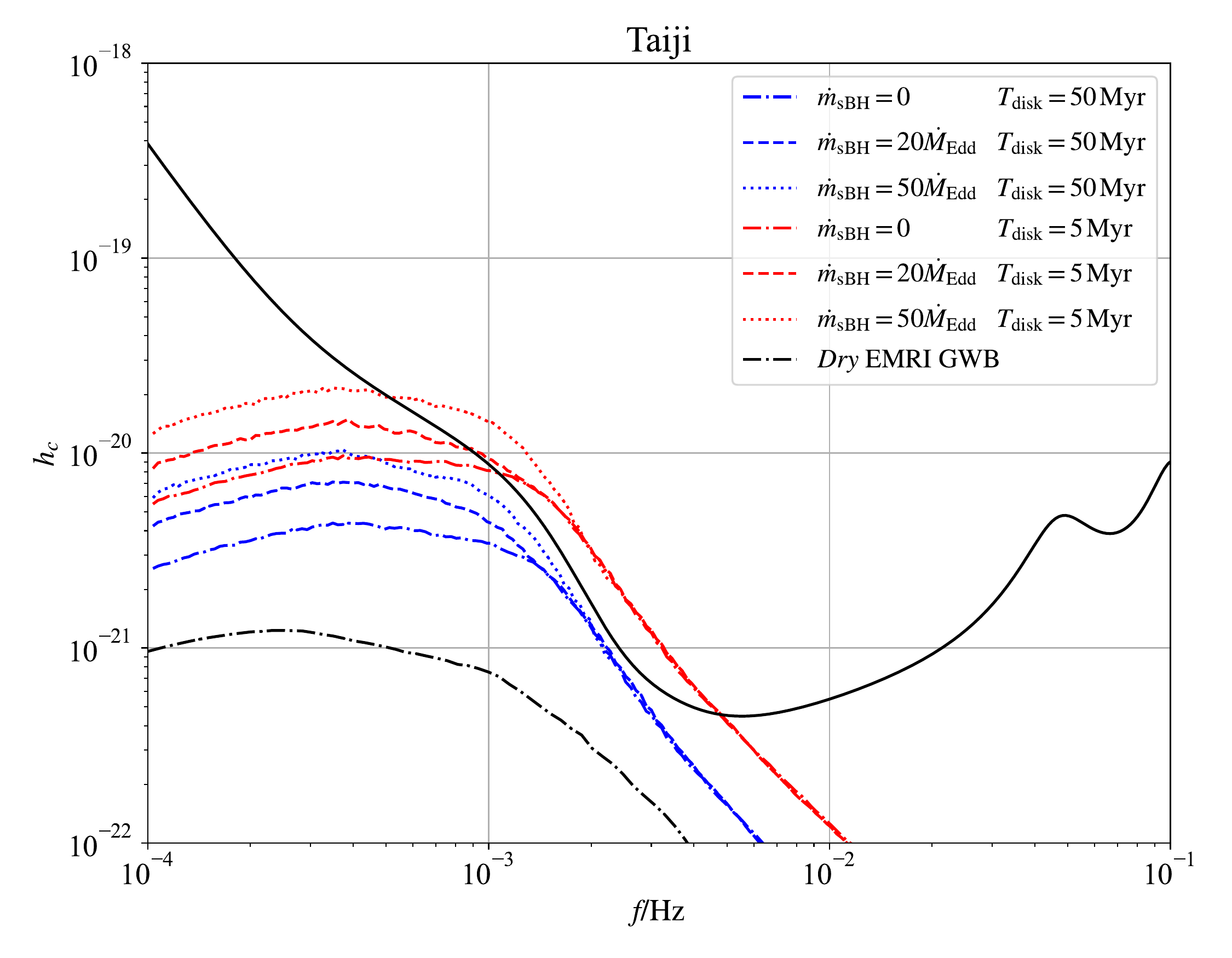}
\includegraphics[width=\columnwidth]{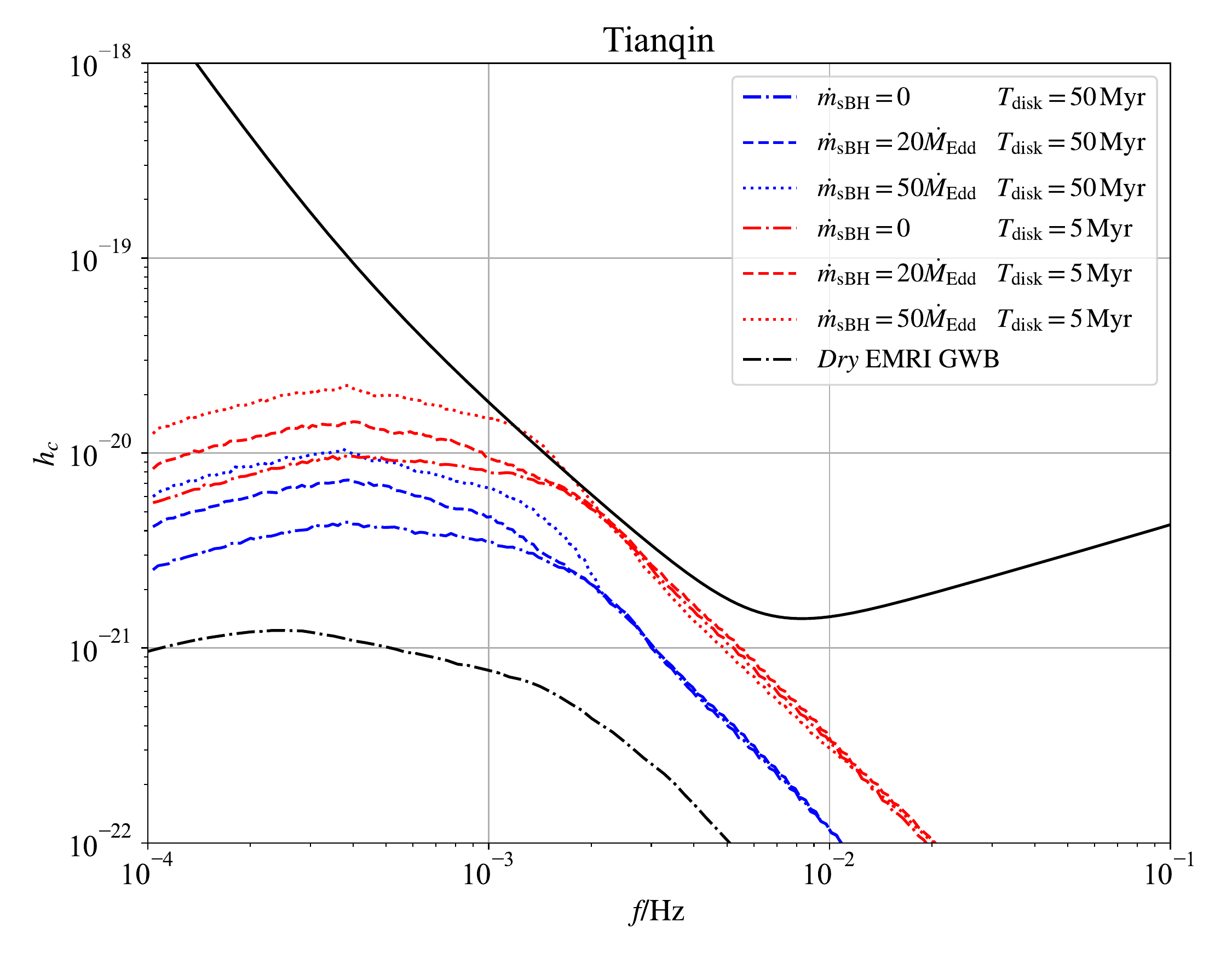}

\caption{GW background of wet and dry EMRI. The solid black lines are the sensitivity curves of LISA/Taiji/Tianqin. The dash-dotted, dashed and dotted lines show the background with accretion rate $\{0,\,20,\,50\}\dot{M}_{\rm Edd}$ for sBHs, respectively, where the disk time of $5\,\rm Myr $ (red) and $50\,\rm Myr$ (blue) are assumed. The black dot-dashed line represents GWB from the dry EMRIs.}
\label{fig:GWB}
\end{figure*}

%\frac{dN}{dz\, dM_\bullet\, de_p\, d\ln f_{\rm orb}}
The characteristic strain spectrum of the stochastic gravitational wave background generated by the EMRI sources can be expressed as \citep[see][ for in detail]{Bonetti2020}:
%\begin{widetext}
\begin{equation}
\begin{split}
&h_{c,\rm dry}^2(f)=\sum\limits_{n=n_{\rm min}}\limits^{n_{\rm max}}\int {\rm d}z\, {\rm d}\,\log M_\bullet \,{\rm d}e_p\\
&\left\{
   \begin{aligned}  
    \left[n(z,M_\bullet,e_p,f_{\rm orb}) h_{n}^2(f)\frac{f^2}{\dot{f}}\frac{1}{fT_{\rm obs}}\right]_{f_{\rm orb}=\frac{f(1+z)}{n}}, \quad T_{\rm obs}>f/\dot{f} , \\
    \left[ n(z,M_\bullet,e_p,f_{\rm orb}) h_{n}^2(f)\right]_{f_{\rm orb}=\frac{f(1+z)}{n}}, \quad T_{\rm obs}<f/\dot{f},
    \end{aligned}
\right.
\end{split}
\end{equation}
%\end{widetext}
where the limits of the harmonic index $n_{\rm min}$, $n_{\rm max}$ are
\be
n_{\rm min}=\frac{f(1+z)}{f_{\rm orb}(t=T_{\rm obs})},\quad n_{\rm max}=\frac{f(1+z)}{f_{\rm orb}(t=0)},
\ee
with $z$ the redshift of the galaxy and $f_{\rm orb}(t)$ is the orbital frequency of the EMRI binary system at time $t$. The strain of the $n$-th harmonic component $h_n$ is given by  \citep[e.g.,][]{Finn2000}:
\be
h_n^2=\frac{G\dot{E}_n}{c^3\pi^2d^2f_n^2}.
\ee
\\
For wet EMRI, the characteristic strain $h^2_{c,\rm wet}$ has the same form as the above equation by setting $n=n_{\rm min}=n_{\rm max}=2$, since the EMRI orbit is mostly circularised. The \emph{dry/wet} EMRI rate in the comoving volume of the universe $V_c(z)$ up to redshift $z$ is:
\be \label{eq:catalog}
\begin{split}
&n(z,M_\bullet,e_p,f_{\rm orb})\equiv\frac{{\rm d}^4 N_{\rm dry/wet}}{ {\rm d}z\, {\rm d} \log M_\bullet\, {\rm d} e_p\, {\rm d} \ln f_{\rm orb} }=\frac{{\rm d^2}N_{\rm dry/wet}}{{\rm d} e_p\,{\rm d}\ln f_{\rm orb}}\times\\
&\frac{1}{1+z}\frac{{\rm d}N_\bullet}{{\rm d}\log M_\bullet}\frac{{\rm d}V_c(z)}{{\rm d}z} C_{\rm cusp}(M_\bullet,z)f_{\rm dry/wet}\bar\Gamma_{\rm dry/wet}(M_\bullet)T_{\rm obs},\\
\end{split}
\ee
where the factor $1/(1+z)$ arises from the cosmological redshift, the fraction of AGN $f_{\rm wet}$ is assumed as $1\%$ \citep[$f_{\rm wet}\sim 1-10\%$, see][]{Paul2012,Macuga2019} and $f_{\rm dry}\equiv 1$, $C_{\rm cusp}(M_\bullet,z)$ is the fraction of MBHs in the stellar cusps which are supposed to be evacuated during mergers of binary MBHs and regrow afterward \citep[][]{Babak2017}, and $dN_\bullet/dM_\bullet$ is the redshift-independent MBH mass functions \citep[][]{Barausse2012},
\be
\frac{{\rm d}N_\bullet}{{\rm d}\log M_\bullet}=0.005\left(\frac{M_\bullet}{3\times 10^6M_\odot} \right)^{-0.3} \rm Mpc^{-3}.
\ee
The $\bar\Gamma_{\rm dry/wet}(M_\bullet)$ has been obtained in the previous section.

To calculate the EMRI GWB, we construct a population with $\{M_\bullet,m_{\rm sBH},z\}$ following the procedure in \citet{Bonetti2020}. For the dry plunge events, the eccentricity at the last stable orbit $e_p$ is sampled from a flat distribution within $e_p\in [0, 0.2]$. For the wet plunge events, we set $e_p=0$ due to the quick circularisation on the disk plane.  Note that not only low SNR ($<20$) plunging systems, but also EMRIs in the LISA band which are still tens or even hundreds of years far from the final plunge contribute to the GWBs. Thus, we randomly sample $N_{\rm back}={\rm int}(T_{\rm back}/\rm yr)$ points in the range $t_{\rm }\in [0, T_{\rm back}]$, which represents $N_{\rm back}$ different EMRIs with different starting time $t_{\rm ini}$. Here the $T_{\rm back}^{\rm dry}$ is assumed by \citep[][]{Bonetti2020}
\be
T_{\rm back}^{\rm dry}=20\left(\frac{M_\bullet}{10^4M_\odot}\right) \,\rm yr,
\ee
For wet EMRIs with the assumption that all orbits are circular, the maximum backward time is set by the observation frequency window of the spaceborne GW detectors. Suppose we set the lower limit of the space-borne detector to be $10^{-4}$\,Hz, then we have $T_{\rm back}^{\rm wet}=T^{\rm wet}_{\rm back}[2f_{\rm orb}/(1+z)=10^{-4}\rm Hz]$. The next step is we integrate the orbital evolution backward in time for $t_{\rm back}\sim[0, T_{\rm back}]$ for each of $N_{\rm back}$ EMRI samples following \citet{peters1964}, i.e., 
\be
\{M_\bullet,m_{\rm sBH},z,e_p,f_p\}\stackrel{t_{\rm back}}{\longrightarrow} \{M_\bullet,m_{\rm sBH},z,e(t=0),f_{\rm orb}(t=0)\},
\ee
where $\{e_p,f_p\}$ and $\{e(t=0),f_{\rm orb}(t=0)\}$ are eccentricity and orbital frequency at plunge time and starting time, respectively. Then we evolve these new EMRIs $\{M_\bullet,m_{\rm sBH},z,e(t=0),f_{\rm orb}(t=0)\}$ forward in time for the observation time $T_{\rm obs}$ of a space GW detection mission, and obtain $\{M_\bullet,m_{\rm sBH},z,e(t=T_{\rm obs}),f_{\rm orb}(t=T_{\rm obs})\}$. It should be noted that $e(t=0)=e(t=T_{\rm obs})=0$ for wet EMRI due to the strong circularisation.

\begin{table}
    \centering
	\fontsize{8}{11}\selectfont    %{字体尺寸}{行距}
	\begin{threeparttable}
	\begin{tabular}{cccccc}
		\toprule
	    \multirow{2}{*}{Model}&\multirow{2}{*}{$T_{\rm disk}/\rm Myr$}&\multirow{2}{*}{$\dot{m}_{\rm sBH}/\dot{M}_{\rm Edd}$}&\multicolumn{3}{c}{$\rm (S/N)_{\rm GWB}$} \\
		 \cmidrule(lr){4-6}
		  & & &LISA&Tianqin&Taiji\\
		\cmidrule{1-6}
		\multirow{6}{*}{wet}
		           &$5$ & 0 & 1526 &183& 2790 \\
		           &$5$ & 20&1698 & 204 & 3126 \\
		           &$5$ & 50 &1879 & 264 & 3261 \\
		           &$50$ & 0 &236 & 29 & 424 \\
		           &$50$ & 20 &255 & 32 & 456 \\
		           &$50$ & 50 &292 & 43 & 484 \\
		 \cmidrule{1-6}
		 dry&$--$ & $--$ & 18 &3.7 & 28 \\
		\bottomrule
	\end{tabular}\vspace{0cm}
	\end{threeparttable}
\caption{SNR of GWB for mission duration $T_{\rm obs}=4\rm yr$. The first column (\emph{Wet/Dry}) is the EMRI model. The second column is the AGN disk lifetime. The third column is the accretion rate of sBH in the stable region of the disk. The fourth column is the SNR of GWB for LISA/Tianqin/Taiji. \label{tab:SNR}}
\end{table}

Our result of the GWB in space-borne GW detector with $M_\bullet \in [10^4,\, 10^7]\,M_\odot$ contributed by the EMRIs in the universe (out to $z = 4.5$) is presented in Fig.~\ref{fig:GWB}, where we compare the GWB generated by the dry EMRIs and that generated by the wet EMRIs with different sBH accretion rate $\dot{m}_{\rm sBH}$ and the disk lifetime $T_{\rm disk}$. The typical mission duration of the space-borne GW detectors is assumed to be 4 years, i.e, $T_{\rm obs}=4\,\rm yr$. It should be noted that the setting of the SNR threshold of a single source will have a substantial impact on the GWB \citep[see][]{Bonetti2020} and we have subtracted GWs of individual sources with $\rm S/N> 20$. The accretion of the sBH mostly affects the GWB spectrum at the sub-millihertz frequency region.

The observational effects of GWB contributed by the EMRIs have two different aspects. Firstly, this GWB could be a foreground noise that degrades the sensitivity for detecting some resolvable GW sources, in particular to the detectors such as LISA and Taiji when the accretion rate of sBH is relatively high as shown in Fig.~\ref{fig:GWB}. In this case, this GWB may be considered as unresolved background noise and the effective noise power spectral density can be estimated as $S_{\rm eff}(f)=S_{\rm noise}(f)+S_{\rm GWB}(f)$, where is given by $S_{\rm GWB}(f)=h_{c,\rm wet}^2(f)/f$. For example, for an EMRI system with $M_\bullet=10^6\,M_\odot$, $m_{\rm sBH}=10\,M_\odot$, eccentricity $e=0$ and redshift $z=0.2$. The accumulated SNR of the combined source after four years of observation to the final plunge is ${\rm SNR}=(36,16,49)$ for\, (LISA, Tianqin, Taiji) for the model with the strongest GWB\,($\dot{m}_{\rm sBH}=50\dot{M}_{\rm Edd}$ and $T_{\rm disk}=5\,\rm Myr$), while we have ${\rm SNR}=(68,20,97)$ when the GWB is absented. Secondly, this GWB itself can be a targeted source carrying useful information about the galaxy centre environment. In this case, the detectability of the GWB is assessed by computing the associated power signal-to-noise ratio $({\rm S}/{\rm N})_{\rm dry/wet}$ through \citep[][]{Thrane2013,Sesana2016}
\be
({\rm S}/{\rm N})_{\rm dry/wet}^2=T_{\rm obs}\int \gamma (f)\frac{h_{c,\rm dry/wet}^4(f)}{f^2 S_{\rm noise}^2(f)}df
\ee
where $S_{\rm noise}(f)$ is the power spectral density of future space-borne GW detectors and the function $\gamma (f)$ is assumed to be approximately constant and equal to unity \citep[see][]{Thrane2013}. We list the SNR of dry and wet GWB in Table\,\ref{tab:SNR}. All the models of wet EMRI GWB listed in Table~\ref{tab:SNR} have a significant signal-to-noise ratio and can be easily detected within 4 years of observation.

\section{Discussion and Conclusions}\label{sec:8}
In this work, we investigate the accretion effect on the distribution of stars and stellar mass black holes in the nuclear region surrounding the MBHs, by self-consistently evolving the extended Fokker-Planck equations. The time evolution of the distribution functions of the stellar mass black holes and stars are carefully analysed. We also study how the rate of the extreme-mass-ratio-inspiral events and their associated gravitational wave background is affected by the accretion effect. The observational consequences have also been presented in this work. (1) The accretion effect shifts the sBH distribution function over the mass to the heavier range, which will have effects on the future observations of stellar mass gravitational wave sources. Our analysis and estimation show that the mergers of sBHs, in particular those heavy sBHs with masses larger than 20 solar mass in the AGN accretion disk, could be an important component of the mass-distribution of the BBH merger events observed by the future 3rd generation ground-based gravitational wave detectors such as ET \citep{Maggiore2020}, Cosmic Explorer \citep[][]{Abbott2017}. (2) On the aspect of space-borne gravitational wave detection, our work is relevant to the 
gravitational wave emitted from the EMRI systems. We found that the accretion effect has the possibility of degrading the detection of resolvable EMRI sources by a factor of $1.5\sim 2$ via forming a stochastic gravitational wave foreground. Moreover, for the EMRI-generated gravitational wave background as the target source for future space-borne detectors, the accretion effect has only a minor effect on the signal-to-noise ratio.

%\R{In this work, we extended the Fokker-Planck equation to the mass-varying scenario and evolved the distribution function of stars/sBHs around a centre MBH. We've concluded that considering the mass accretion effect of stars/sBHs will have a non-negligible impact on the detection of GW on the ground-based and space-borne GW detectors.  }
However, the results presented in this work are still primitive. Many assumptions have been imposed in our modeling due to the complexity of the nuclear environment and the relevant physical processes, which are worth further discussions as follows.

Firstly, the damping timescale of orbiters' inclination is much shorter than the migration timescale only for low-inclination orbiters. For orbiters with high inclination angles, the dynamical friction (or the drag force) will dominate the damping process of the inclination angle, the timescale of which could be comparable to or longer than the migration time in disk \citep[e.g.,][]{Cresswell2008,Rein2012,Arzamasskiy2018MNRAS,Zhu2019MNRAS,Nasim2022}.
In this case, the low-inclination orbiters \citep[with inclination angle $i<\pi/4$, see][]{Nasim2020,Fabj2020,Nasim2022} with shorter capture timescale will participate in the interaction with the disk\,(e.g. dynamic capture and scattering process) while the high-inclination ones with longer capture timescale remain in the cluster. Thus, the rates of replenishment of disk orbiters in this work are an upper limit since a significant fraction of orbiters with $i > \pi/4$ take significantly longer to be captured.
In our calculation, the time average capture rate is obtained in the parameter $\mu_{s/\rm sBH}$. One can phenomenologically treat the effects of different capture rates for these low- and high-inclination orbiters by adjusting the parameter $\mu_{s/\rm sBH}$.
%the capture rate $\mu_s$ is determined by an equilibrium between diffusion by two-body encounters and energy loss by the dissipative force from the disk \citep[][]{Vilkoviskij2002,Kennedy2016,Panamarev2018}. 
It should be noted that, for numerical simplification, we still adopt the phenomenological treatment \citep{PanZhen2021} for the disk capture process rather than calculate it from the first principle \citep{PanZhen2022}. The time-averaged wet EMRI rates by the phenomenological treatment are generally consistent with that from the first principle \citep[see][]{PanZhen2022}. 
%What confirms us to use the phenomenological treatment is that the time-averaged wet EMRI rates are generally consistent with that from the first principle \citep[see][]{PanZhen2022}.  

Secondly, we did not include the migration process driven by dynamical friction in our calculation, which is also called type-0 migration \citep[e.g.,][]{Syer1991MNRAS,Vilkoviskij2002,Nasim2022,Paardekooper2022}. The drag force is non-negligible when the mass ratio is small ($q\equiv m_{\rm sBH}/M_\bullet<10^{-7}$) and the gravitational interaction does not dominate \citep[see][for reviews]{Kley2012,Paardekooper2022}. For the EMRI system we focus on ($q\sim 10^{-6}-10^{-3}$), the migration timescale and the capture timescale driven by density wave are several orders of magnitude shorter than that by dynamical friction for the low inclination orbiters \citep[with orbital inclination angle $i<\pi/4$, see][]{Cresswell2008,Rein2012,Arzamasskiy2018MNRAS,Fabj2020,Nasim2022}. However, for the higher inclination and retrograde orbiters\,($\pi/4\le i\le\pi$), the dynamical friction will dominate the migration process and the damping of inclination angle, which will affect our results in two ways: (1) It can decrease the disk capture rate $\mu_{s/\rm sBH}$ after the low inclination orbiters are depleted. One can relax this problem by adjusting the $\mu_{s/\rm sBH}$. (2) It also affects the migration timescale of retrograde orbiters in the disk ($i=\pi$). The migration timescale of retrograde orbiters is about $5-10$ times longer than that of prograde orbiters in disk ($i=0$) \citep[e.g.,][]{Secunda2021ApJ,Nasim2022}, which could decrease the wet EMRI rate depending on the fraction of retrograde orbiters in the disk. However, in the disk, the fraction of retrograde orbiters could be much smaller than that of prograde orbiters because the stars/sBHs are mainly captured from the nuclear star cluster in active stage. For example, the simulation result in \citet{Arzamasskiy2018MNRAS} showed that the capture timescale of low inclination orbiters in the prograde orbits ($i<\pi/4$) driven by density wave is 1-2 orders of magnitude shorter than that of retrograde orbiters ($\pi/2<i<\pi$) driven by dynamical friction. 

Thirdly, we also did not include the outward migration in our calculation \citep[see][]{Masset2003ApJ,Pepli2008MNRASa,Pepli2008c,McKernan2011}. The outward migration could occur when the star/sBH perturbs the disk, but is not strong enough to open a gap. The mass ratio $q$ is about $10^{-5}-10^{-4}$ for outward migration \citep[][]{McKernan2011} depending on the disk aspect ratio $h(r)$, surface density $\Sigma(r)$ and the viscous coefficient $\alpha$, which generally need to be calculated by numerical simulation \citep[][]{Masset2003ApJ,Pepli2008MNRASa,Pepli2008c}. The outward migration could decrease the EMRI rate due to its fast runaway migration. However, sustaining this rapid mode of migration has proved to be very difficult, which strongly depends on the specific state of gas flow in the star/sBH’s vicinity.  For example, it needs a stringent condition such as a very steep positive surface density gradients in the disk \citep[][]{Pepli2008MNRASa,Pepli2008c,Kley2012}.
%However, the specific conditions of outward migration are somewhat stringent: (1) In the simulations of \citet{Masset2003ApJ} and \citet{Pepli2008MNRASa}, the spontaneous outward migration is not found in disk with smooth profiles. It means the outward migration needs very steep positive surface density gradients in the disk \citep[][]{Kley2012}. 
%(2) The outward migration strongly depends on the gap opening conditions. The type-II migration will be triggered when the gap is too deep, while the type-I migration will be triggered when the gap is too shallow. It means that the mass ratio $q$ for outward migration could be limited to a small range  \citep[][]{Masset2003ApJ,Kley2012}. Due to the above reasons, only a fraction of the EMRI systems will experience outward migration, which could have a small influence on the overall GWB of EMRI. 
Due to the strigent condition for the occurrence of the outward migration and the complexity of this issue, we temporarily did not consider the impact of outward migration in this work.

Fourthly, we adopted a fiducial model in this work, such as the total relative abundance of sBHs $\varphi=0.001$ and the fraction of AGN $f_{\rm wet}=1\%$. As shown in Fig.\,\ref{fig:GWB}, the GWB of wet EMRI has a non-negligible impact on space-borne GW detectors, especially LISA/Taiji. It should be noted that the characteristic strain of GWB $h_{c,\rm wet}^2\propto \varphi f_{\rm wet}$, where wet GWB might have a stronger effect on the sensitivity curve with more sBHs and AGNs. In addition, the duty cycle of AGNs $T_{\rm disk}$ is still unclear. Estimated by the viscous timescale, the duty cycle is proportional to $\sqrt{M_\bullet}$ and is approximately in a range of $10^6\sim 10^8\rm yr$ for $M_\bullet=10^4\sim 10^7 M_\odot$. However, the duty cycle is associated with many physical factors including gas supply and feedback process, etc. Here, we just take a homogeneous distribution of the duty cycle distribution  $T_{\rm disk}=\{5, 50\}\,\rm Myr$ \citep[e.g.,][]{Shulevski2015,Turner2018} in this work.

Fifthly, the accretion process of stars/sBHs in the disk is quite unclear. We assume the accretion of sBHs is about 1-50 $\dot M_{\rm Edd}$ in the stable region and $\dot M_{\rm Edd}$ in the unstable region \citep[][]{Yangxiaohong2014,McKinney2014,PanZhen2021accretion}. For the accretion of stars, we take an empirical accretion model fitted from the universally high abundance ratio of [Fe/Mg] in the inner stable region \citep[][]{Toyouchi2022} and also assume the accretion rate is $\dot{M}_{\rm Edd}$ in the unstable region since the disk could be fragmentation and collapse into clumps. In addition, the evolution of stars in disk is also uncertain. We assume that stars with mass $50\,M_\odot<m_s<150\,M_\odot$ will collapse into sBHs after $5\,\rm Myr$ \citep[][]{Toyouchi2022}. During the active stage\,($< 100\,\rm Myr$), fewer than 100 stars collapse into sBHs approximately, which has a small effect on the EMRI event rate, and mass distribution of sBH. However, in fact, the stars could have a higher accretion rate as shown in \citet{Wangjianmin2021,Matto2021,Dittmann2021,Jermyn2021}, which indicates the possibility that more massive stars will collapse into sBHs.  

Sixthly, the equation~\eqref{eq:binary_converter} and the associated parameters we used in the work is a simplified approximation. The binary mergers in the AGN accretion disk are also a complicated process that triggers many studies \citep[e.g.,][]{Grobner2020,Graham2020,Gerosa2021,Li2022,Li2022arXiv,Samsing2022Nature}. However, there are still many uncertain factors in the disc-binary interaction processes yet to be understood, which can affect our results of the ${\rm d}R/{\rm d}m_{\rm sBH}$.

Seventhly, the binaries of stars/sBHs are not included in our calculation of the Fokker-Planck equations, which could have the following impact on our results. On one hand, the tidal capturing of one component in the binary could contribute to the EMRI rate \citep[][]{Hills1988}, which could be comparable to the EMRI rate driven by the two-body scattering \citep[see][for a review]{Pau2018}. On the other hand, the dynamical effect of binaries (i.e., the three- and four-body encounters between binaries and single stars/binaries) may be important in the evolution of stars/sBHs' distribution function, which generally needs to be calculated by N-body or Monte Carlo simulations \citep[e.g.,][]{Spurzem1996,Giersz2000,Pau2018}. In this way, the EMRI rate may also affected by the dynamical effect of binaries. These effects will be considered in the future work.

\section{Acknowledgements}
Y.\,M. thanks Professor Yanbei Chen for helpful discussions on the LIGO binary black hole detection, and Dr. Ning Jiang for the discussion on the AGN physics.  He also thanks Professors Shun Wang and Jingtao L{\"u} for useful conversations on non-equilibrium statistical physics. Q.\,W. thanks Professors Jianmin Wang and Luis C. Ho for the very useful discussions and comments on the disk stars. Y.\,M. is supported by the university start-up funding provided by Huazhong University of Science and Technology. M.\,W. and Q.\,W. is supported by the NSFC (grants U1931203, 12233007) and the science research grants from the China Manned Space Project (No. CMS-CSST-2021-A06). The authors acknowledge Beijing PARATERA Tech CO.,Ltd. for providing HPC resources that have contributed to the research results reported within this paper. \\

\noindent DATA AVAILABILITY STATEMENT: The data underlying this article will be shared on reasonable request to the corresponding authors.

\appendix
\section{Diffusion and advection coefficients in the Fokker-Planck equation}\label{appendix}
In this section, we extend the calculation of the diffusion and the advection coefficients of two-component (stars and sBHs with single mass) in \citet{PanZhen2021} to N-component (stars and sBHs with multiple mass) cases. We first define a few auxiliary functions:
\be
\begin{split}
&F_0^i(m_i,E,r)=(4\pi)^2 {Gm_i}^2\ln{\Lambda}\int^E_{-\infty}dE'\bar f_i(m_i, E'),\\
&F_1^i(m_i,E,r)=(4\pi)^2 {Gm_i}^2\ln{\Lambda}\int_E^{\phi(r)}dE'\left(\frac{\phi-E'}{\phi-E}\right)^{1/2} \bar f_i(m_i, E'),\\
&F_2^i(m_i,E,r)=(4\pi)^2 {Gm_i}^2\ln{\Lambda}\int_E^{\phi(r)}dE'\left(\frac{\phi-E'}{\phi-E}\right)^{3/2}\bar f_i(m_i, E'),
\end{split}
\ee
where $i=\{\rm star, sBH\}$, $\ln \Lambda$ the Coulomb’s logarithm which take as $\ln \Lambda = 10$, and 
\be
\bar f_i(m_i,E')=\int_0^1 d\mathscr{R} f_i(E,\mathscr{R},m_i).
\ee
With these auxiliary functions, the coefficients are written as
\be
\begin{split}
&D_{EE}^i(E,\mathscr{R},m_i)=\frac{8\pi^2}{3}J_c^2\int dm_i' \int^{r_+}_{r_-}\frac{dr}{v_r}v^2\left[F_0^i(m_i',E,r)+F_2^i(m_i',E,r)\right],\\
&\qquad\qquad\qquad\quad+(i\leftrightarrow j),\\
&D_E^i(E,\mathscr{R},m_i)=-8\pi^2J_c^2\int dm_i'\frac{m_i}{m_i'} \int_{r_-}^{r_+}\frac{dr}{v_r}  F_1^i(m_i',E,r)+\frac{m_i}{m_j}\times (i\leftrightarrow j),\\
&D_{E\mathscr{R}}^i(E,\mathscr{R},m_i)=\\
&\frac{16\pi^2}{3}J^2\int dm_i'\int_{r_-}^{r_+}\frac{dr}{v_r}\left(\frac{v^2}{v_c^2-1}\right) [F_0^i(m_i',E,r)+F_2^i(m_i',E,r)]\\
&\qquad\qquad\qquad\quad+(i\leftrightarrow j),\\
&D_{\mathscr{RR}}^i(E,\mathscr{R},m_i)=\\
&\frac{16\pi^2}{3}\mathscr{R}\int dm_i'\int_{r_-}^{r_+}\frac{dr}{v_r}\left\{2\frac{r^2}{v^2}\left[v_t^2\left(\frac{v^2}{v_c^2}-1\right)^2+v_r^2\right]F_0^i(m_i',E,r)\right.\\
&\left. +3\frac{r^2}{v^2}v_r^2 F_1^i(m_i',E,r)+\frac{r^2}{v^2}\left[2v_t^2\left(\frac{v^2}{v_c^2}-1\right)^2-v_r^2\right]F_2^i(m_i',E,r)\right\}\\
&\qquad\qquad\qquad\quad+(i\leftrightarrow j),\\
&D_\mathscr{R}^i(E,\mathscr{R},m_i)=-16\pi^2 \mathscr{R} r_c^2\int dm_i' \frac{m_i}{m_i'}\int_{r_-}^{r_+} \frac{dr}{v_r}\left(1-\frac{v_c^2}{v^2}\right)F_1^i(m_i',E,r)\\
&\qquad\qquad\qquad\quad+\frac{m_i}{m_j}\times (i\leftrightarrow j),
\end{split}
\ee
where $j= \{{\rm star, sBH}\}$, $i\not= j$, and $v_t = J/r$ is the tangential velocity.

\bibliographystyle{mnras}
\bibliography{ref}
\bsp	% typesetting comment
\label{lastpage}

\end{document}